\documentclass[epsf,preprint,aps]{revtex4}
\usepackage[dvips]{epsfig}
\usepackage{graphicx}
\usepackage{xcolor}
\usepackage{bm}
\usepackage{amsmath}

\date{\today}

\begin{document}

\title{Planar channeling and quasichanneling oscillations in a bent crystal}

\author{A.I. Sytov$^{1,2}$}
\email{alex_sytov@mail.ru}
\author{V. Guidi$^{1}$, V.V. Tikhomirov$^{2}$, E. Bagli$^{1}$, L. Bandiera$^{1}$, G. Germogli$^{1}$, A. Mazzolari$^{1}$}
\address{$^1$ INFN Sezione di Ferrara, Dipartimento di Fisica e Scienze della Terra, Universit$\grave{a}$ di Ferrara, Via Saragat 1, 44100 Ferrara, Italy.}
\address{$^2$ Research Institute for Nuclear Problems, Belarusian State University, Bobruiskaya str., 11, 220030, Minsk, Belarus.}

\begin{abstract}
Particles passing through a crystal under planar channeling experience transverse oscillations in their motion. As channeled particles approach the atomic planes of a crystal, they are likely to be dechanneled. This effect was used in ion-beam analysis with MeV energy. We studied this effect in a bent crystal for positive and negative particles within a wide range of energies in sight of application of such crystals at accelerators. We found the conditions for the appearance or not of channeling oscillations. Indeed a new kind of oscillations, strictly related to the motion of over-barrier particles, i.e. quasichanneling particles, has been predicted. Such oscillations, named planar quasichanneling oscillations, possess a different nature than channeling oscillations. Through computer simulation, we studied this effect and provided a theoretical interpretation for them. We show that channeling oscillations can be observed only for positive particles while quasichanneling oscillations can exist for particles with either sign. The conditions for experimental observation of channeling and quasichanneling oscillations at existing accelerators with available crystal has been found and optimized.
\end{abstract}
\maketitle

\section{Introduction}

\textit{Channeling} is a coherent effect of penetration of charged particles in a crystal almost parallel to its axes or planes. Charged particles under channeling conditions move in the electric field of atoms, which builds up the averaged transverse interplanar potential and electric field. This concept, called the \textit{continuum potential}, was proposed by J. Lindhard \cite{Lindhard}, who developed the theory of the channeling effect. In the following we consider only motion along crystal planes called \textit{planar} channeling.

The interplanar electric field induces harmonic-like transverse oscillations. These oscillations are called \textit{planar channeling oscillations}, which correspond to an \textit{under-barrier} motion along the crystal planes. The planar oscillation length can be estimated using harmonic approximation:
\begin{equation}
\label{1}
\lambda=\pi d_{0}\sqrt{\frac{pv}{2U_{0}}},
\end{equation}
where $d_{0}$ is the interplanar distance, $p$ and $v$ the particle momentum and velocity respectively, $U_{0}$ the potential well height for a straight crystal.

For positive particles, the oscillation length is nearly the same for the particles with the same energy. This gives rise to \textit{phase correlation} of different trajectories. Depending on the difference in the oscillation lengths, such correlation can be conserved for several or, at certain conditions, even several tens of oscillations.

Phase correlation has already been used in two circumstances. The first one is the so-called mirroring \cite{TT,LEUA9}, i.e., the charged particle reflection from crystal planes in a straight crystal of the length of a half channeling oscillation. The effect of mirroring of 400 GeV/c protons, recently observed at the CERN SPS \cite{LEUA9}, can be applied to particle deflection at future accelerators. It is also possible to observe the oscillations of over-barrier particles in the same thin crystal. Such oscillations are described below. The second can be realized in making a narrow plane cut perpendicularly to the crystal planes, resulting in an increase in channeling efficiency up to 99\% \cite{Tikh1,GuidiTikh}. The idea consists in focusing the particles in the cut to the centers of interplanar channels and their consequent recapture under channeling mode, far away from the crystal planes.

The phase correlation of different trajectories is the main condition for the observation of planar channeling oscillations in the angular distribution behind the crystal. If different trajectories are well correlated in their oscillations, they will synchronously approach the crystal planes. The probability of either Coulomb or nuclear interaction causing an escape from the channeling mode, the so-called dechanneling, is the highest as the particle becomes closer to the planes. Therefore, the distribution of penetration depth of particles in a crystal under channeling mode will possess a periodic-like structure of peaks and deeps. The distance between them will be proportional to the channeling oscillation length.

Planar channeling oscillations at low energies in backscattering were predicted by J.H. Barrett \cite{Ba71,Ba79} in simulations. Later they were observed in several experiments \cite{Abel1,Abel2,Abel3,Kauf,PCO1,PCO2,Berec} with ion beams of the energy of the order of MeV and well described in \cite{Gemmell,Feldman}.

Channeling in a \textit{bent} crystal, as proposed by Tsyganov \cite{Tsyganov}, allowed the deflection of a charged particle beam of the energy from hundreds of MeV up to tens of TeV in many experiments \cite{Tevatron,Tevatron2,UA9,UA92,U70,U702,TikhPRL,TikhPRL2}. At the moment experiments with high energy physics do not take into account the effects of correlations particle trajectories. Indeed, much information can be gained from this knowledge.

\begin{figure}
\resizebox{89mm}{!}{\includegraphics{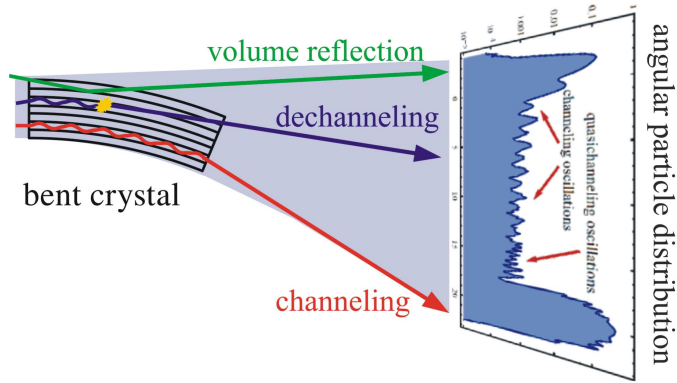}}
 \caption{\label{F000} Sketch of the angular particle distribution behind a bent crystal.}
 \end{figure}

As an example at high energies planar channeling oscillations in crystal are transformed to dechanneling peaks in the deflection angle distribution of the beam passed through the crystal \cite{Sytov}, as shown in Fig. \ref{F000}. This method is applicable only for a bent crystal, allowing to obtain the angular unfolding of the dechanneling process. As we will show below, this possibility can be realized only for positive particles.

The aim of this paper is the prediction of another kind of oscillations in the angular distribution of the particles after interaction with a bent crystal (Fig. \ref{F000}). Differently from planar channeling oscillations, this effect regards the motion of \textit{over-barrier} particles at sufficiently small angles w.r.t. the crystal planes. In analogy to the quasi-channeling motion, such kind of oscillations will be hereinafter named ``\textit{planar quasichanneling oscillations}". We predict the observation of the quasichanneling oscillations for both positive and negative particles. The deflection peak angles are described by the same relation independently of particle charge and energy. We argue that the quasichanneling peak structure is solely determined by the crystal geometry and lattice. We provide simulation results for different energies for different particles of both charge signs interacting with different crystal planes and dimensions and provide theoretical interpretation and comparison with our simulations. We also compare the simulated pictures of channeling and quasichanneling oscillations as well as observe their combination for positive particles. For both cases, we propose an experimental setup as well as an energy scaling of the setup. We finally provide the optimal experimental conditions for both kinds of oscillations for either channeling or volume-reflection orientations.

\section{General background}

\subsection{Channeling in crystals}

\begin{figure}
\resizebox{77mm}{!}{\includegraphics{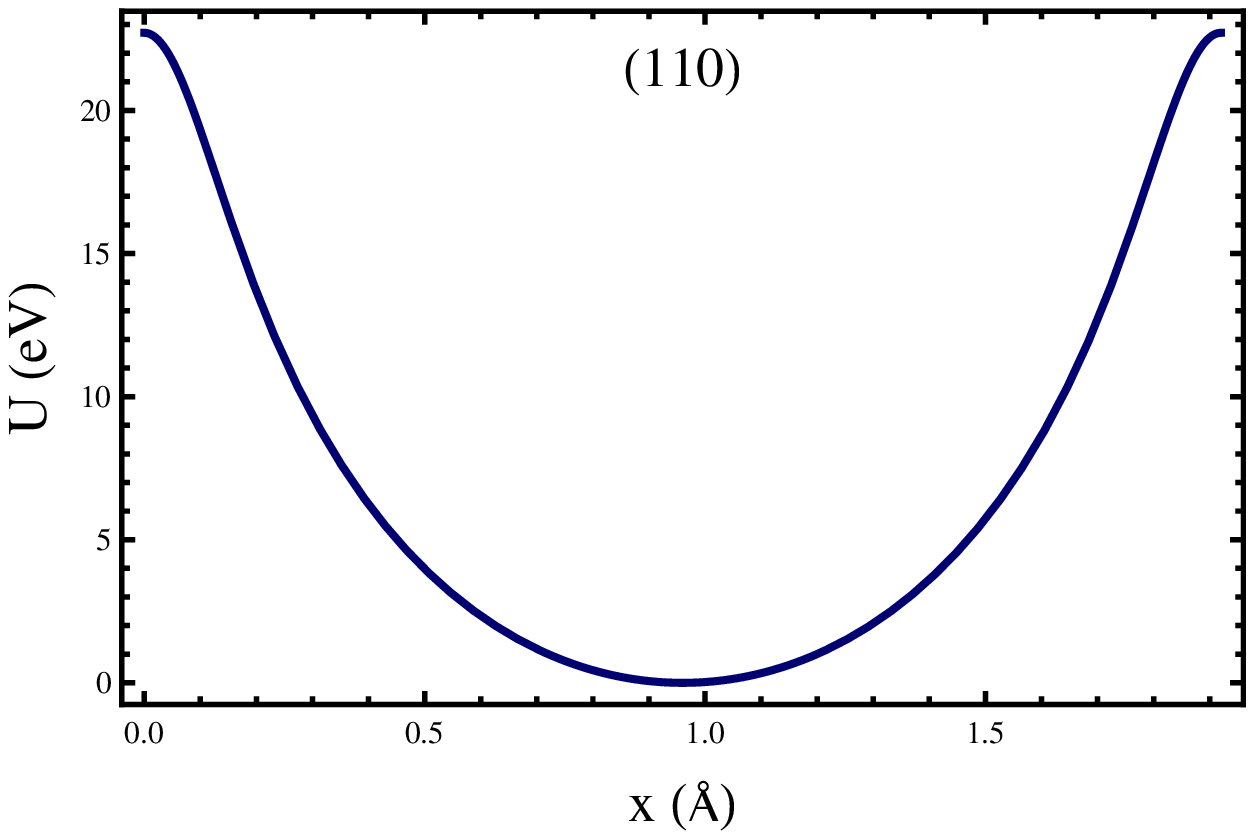}}
\resizebox{77mm}{!}{\includegraphics{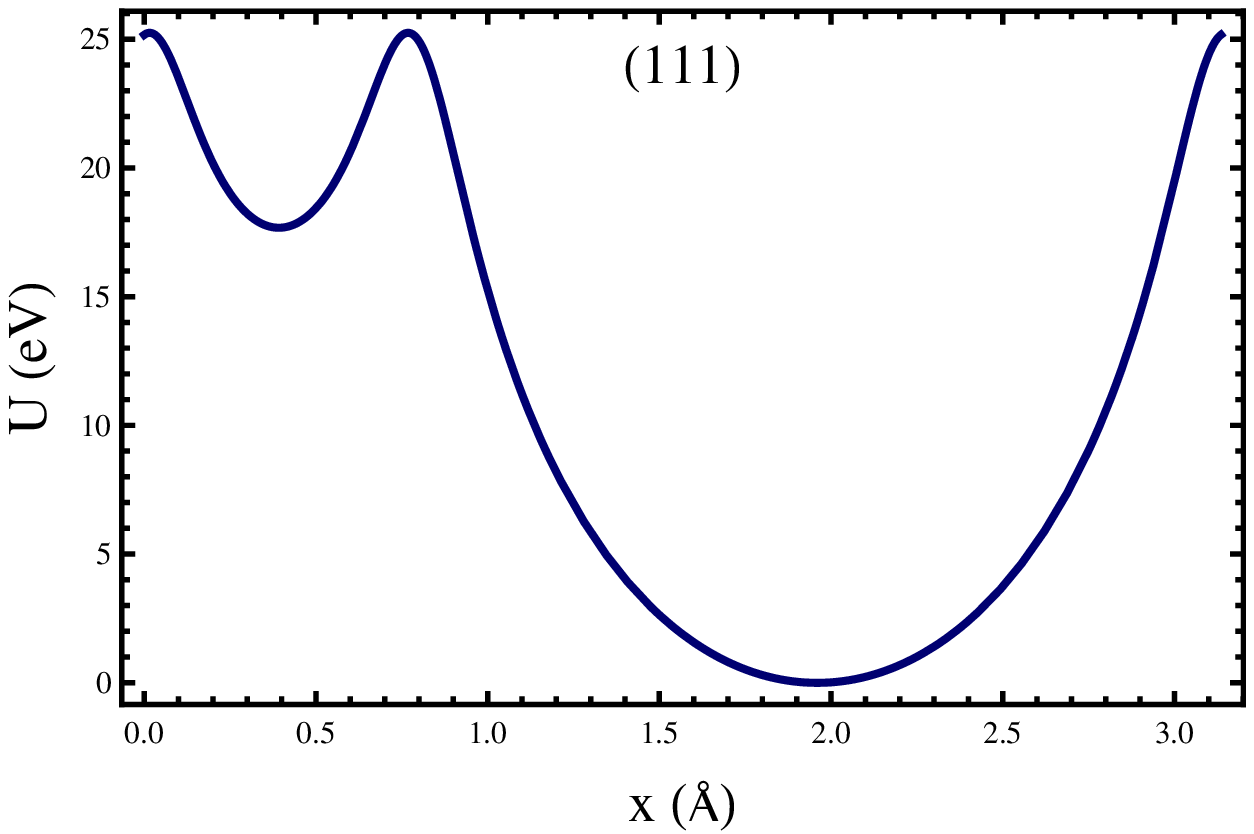}}
 \caption{\label{F00} Interplanar potential well in a straight silicon crystal for (110) (top) and (111) (bottom) planes for positive particles. The crystal planes precisely coincide with the maxima of the potential for (110) or are close to for (111).}
 \end{figure}

As mentioned above, channeling is determined as the effect of penetration of charged particles in a crystal almost parallel to its axes or planes. It is possible to use the continuous approximation of the potential and electric field because of small particle incidence angles w.r.t. to the crystal planes or axes and large longitudinal velocities. In the case of planar channeling, particles will accomplish an oscillatory transverse under-barrier motion (the planar channeling oscillations) in the transverse interplanar potential $U(x)$. This latter is shown in Fig. \ref{F00} under Moli\'{e}re approximation \cite{Appleton,Gemmell,Chu,Biryukov} for both (110) and (111) planes of a straight silicon crystal. This well is for positive particles. For the negative ones it should be taken with opposite sign $-U(x)$, which inverts the picture. The main condition for channeling is the initial angle of a particle $\theta_{in}$ to be less than the critical angle called the Lindhard angle \cite{Lindhard}:
\begin{equation}
\label{31}
\theta_L=\sqrt{\frac{2U_0}{pv}}.
\end{equation}

Effective bent crystal potential is introduced in the comoving reference system and contains a centrifugal term:
\begin{equation}
\label{0}
U_{eff}(x)=U(x)+pvx/R,
\end{equation}
where $R$ is the transverse bending radius of the crystal. This radius should exceed the critical value $R_{cr}$
\begin{equation}
\label{3}
\frac{R}{R_{cr}}=R\frac{U'_{max}}{pv}>1,
\end{equation}
where $U'_{max}$ is the maximal electric field in a straight channel. Otherwise the centrifugal force will exceed the electric one and channeling will not occur any longer.

Channeled particles may escape the channeling mode due to scattering on nuclei and electrons. This is so-called dechanneling effect. The probability of scattering depends on the nuclear and electron densities, which are evidently higher near the crystal planes \cite{Tikh12,Tikh2}. For this reason, the particles with higher amplitudes of the channeling oscillations are likely to dechannel more frequently than that with smaller amplitude.

\subsection{Channeling oscillations}

\begin{figure}
\resizebox{89mm}{!}{\includegraphics{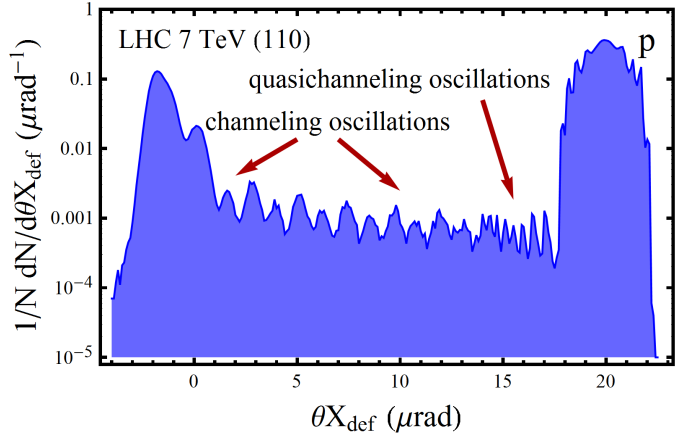}}
 \caption{\label{F0} Angular distribution of the 7 TeV proton beam after interaction with the silicon crystal at the channeling orientation. The simulation layout is: r.m.s. beam angular divergence $\theta_{in}=0.5\mu rad$, crystal length $l_{cr}=2 mm$, bending angle $\theta_b=20\mu rad$, (110) planes.}
 \end{figure}

An example of dechanneling peaks, corresponding to the planar channeling oscillations in the angular distribution behind the crystal, is shown in Fig. \ref{F0} for (110) planes. This result was obtained by our simulations described in the next section.

The origin of the dechanneling peaks consists in a high-phase correlation of trajectories of different particles, dechanneling close to the atomic planes where the nuclear density is high. Note that the number of dechanneling peaks corresponds to the number of particle approaches to a crystal plane where the probability of scattering is high. In other words, the dechanneling peak number in Fig. \ref{F0} is equal to the number of channeling half oscillations.

\begin{figure}
\resizebox{77mm}{!}{\includegraphics{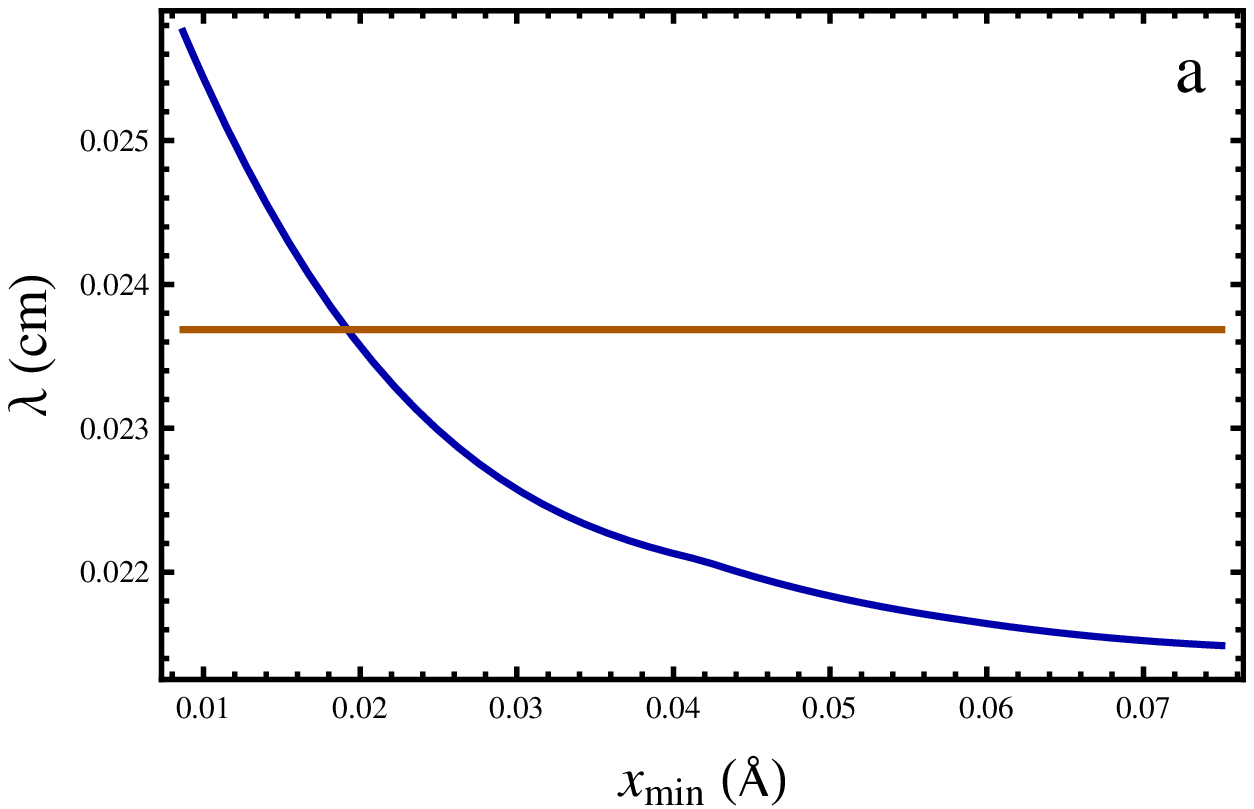}}
\resizebox{77mm}{!}{\includegraphics{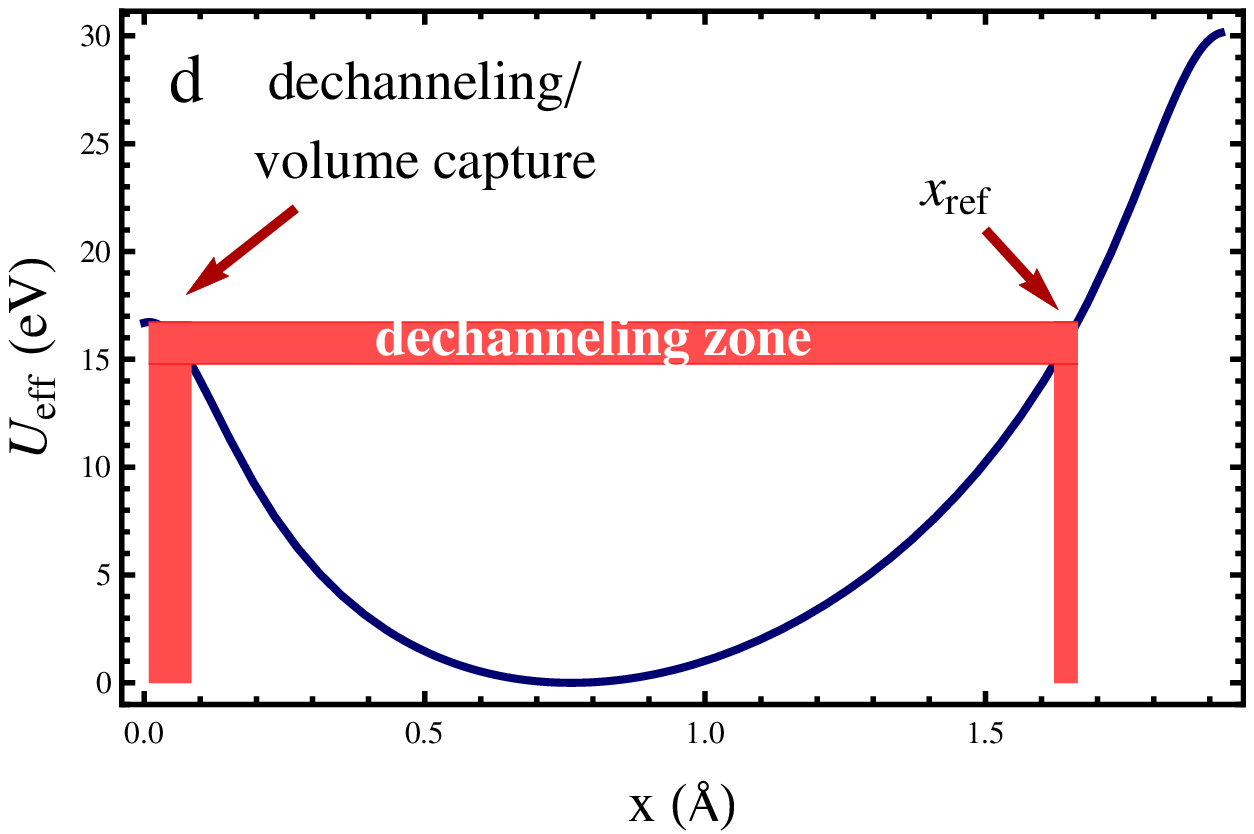}}
\resizebox{77mm}{!}{\includegraphics{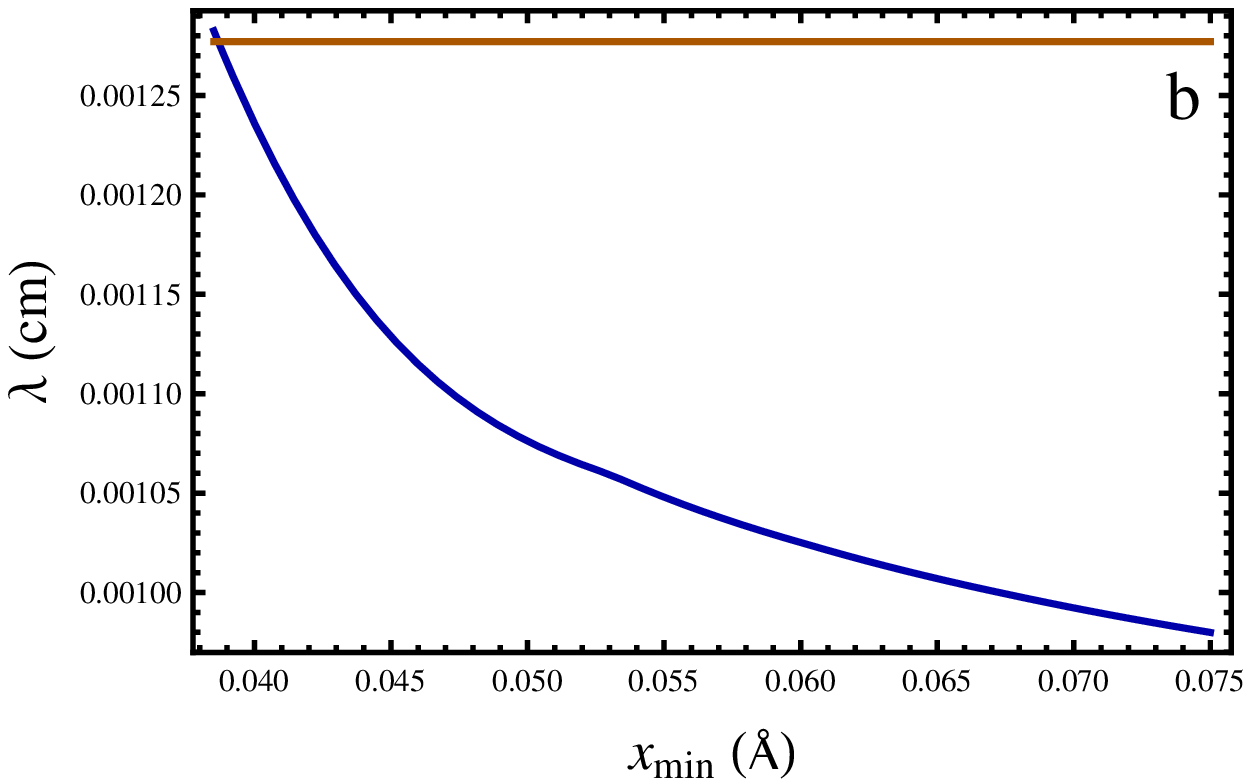}}
\resizebox{77mm}{!}{\includegraphics{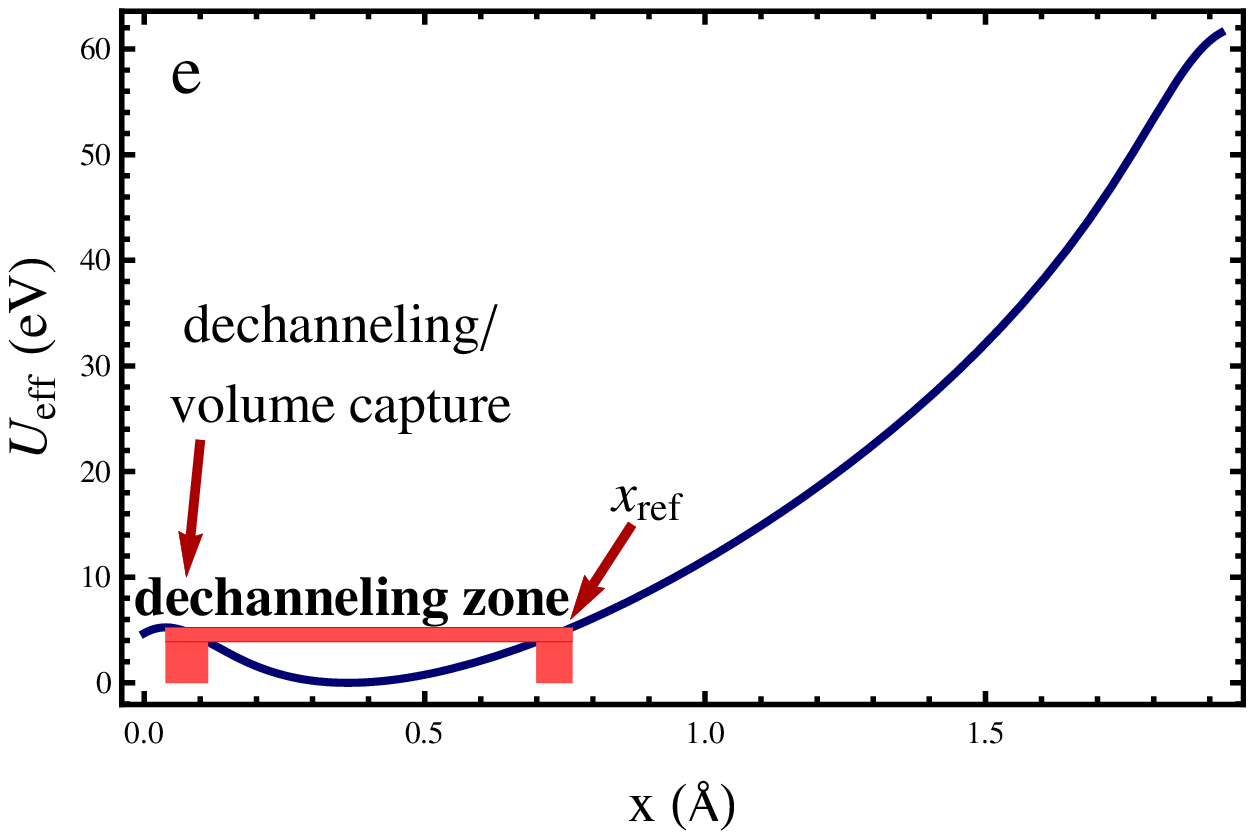}}
\resizebox{77mm}{!}{\includegraphics{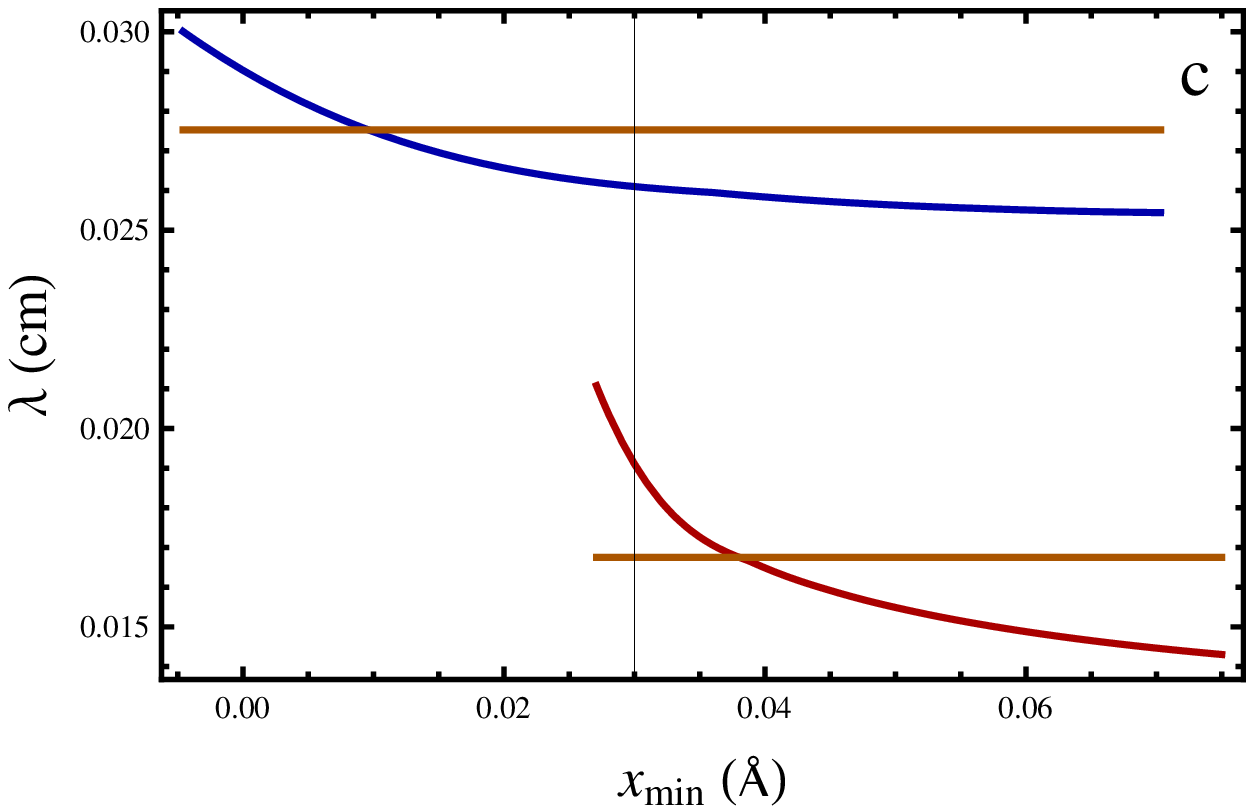}}
\resizebox{77mm}{!}{\includegraphics{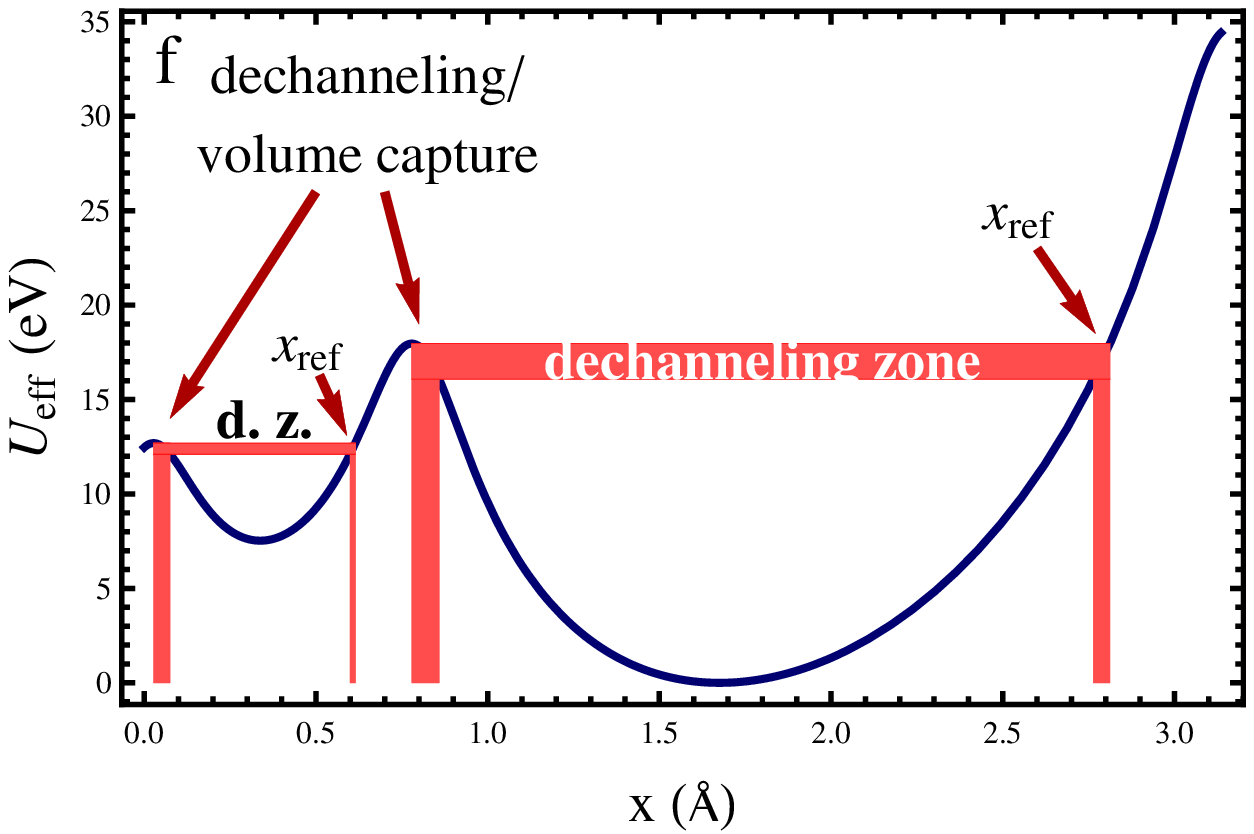}}
 \caption{\label{F2} The channeling oscillation length \textit{vs} the left turning point of the trajectory (left column) and the corresponding potential wells (right column) in a bent crystals:\\
 a,d: 7 TeV protons, $l_{cr}=2 mm$, $\theta_b=20\mu rad$, (110) planes;\\
 b,e: 20.35 GeV positrons, $l_{cr}=0.11 mm$, $\theta_b=1600\mu rad$, (110) planes;\\
 c,f: 7 TeV protons, $l_{cr}=2 mm$, $\theta_b=20\mu rad$, (111) planes. Larger channeling oscillation length corresponds to the wider potential well. \\
 Horizontal lines correspond to the channeling length estimation by (\ref{1}). The left point placed at x=0 indicates the coordinate of the crystal plane.
}
 \end{figure}

The channeling oscillation length can be evaluated directly by integration of the equation of motion. Examples of the dependence of this length on the coordinate of the left turning point of the trajectory are shown in Fig. \ref{F2}. The corresponding potential wells are also drown in Fig. \ref{F2}. These plots represent some cases considered in the next section. Note that the particles dechannel with highest probability only near the lower potential maximum, to which the particles approach closer as shown in Fig. \ref{F2}.

The width of the dechanneling zone is determined by the amplitude of atomic thermal vibrations, which is equal to 0.075${\AA}$ for silicon at the room temperature \cite{Gemmell}. Indeed, it is shown in Fig. \ref{F2} that indeed the channeling oscillation length varies rather weakly in the dechanneling zone and Eq. (\ref{1}) can be applied. Consequently there is a phase correlation of different trajectories for positive particles in the dechanneling zone. Thus, such particles dechannel almost at the same depths.

The decrease of the ratio of the crystal bending radius to its critical value $R/R_{cr}$ reduces the phase correlation of the trajectories, resulting in deterioration of the structure of dechanneling peaks as will be shown below. The length estimated by Eq. \ref{1} becomes a bit overestimated at small radii of curvature (see Fig. \ref{F2}b). This results in a higher number of dechanneling peaks.

The channeling oscillation length in Fig. \ref{F2} is proportional to $\sqrt{pv}$ (like in the formula (\ref{1})) for fixed form of the potential well. Thereby, phase correlation should take place for different energies of positive particles.

\begin{figure}
\resizebox{89mm}{!}{\includegraphics{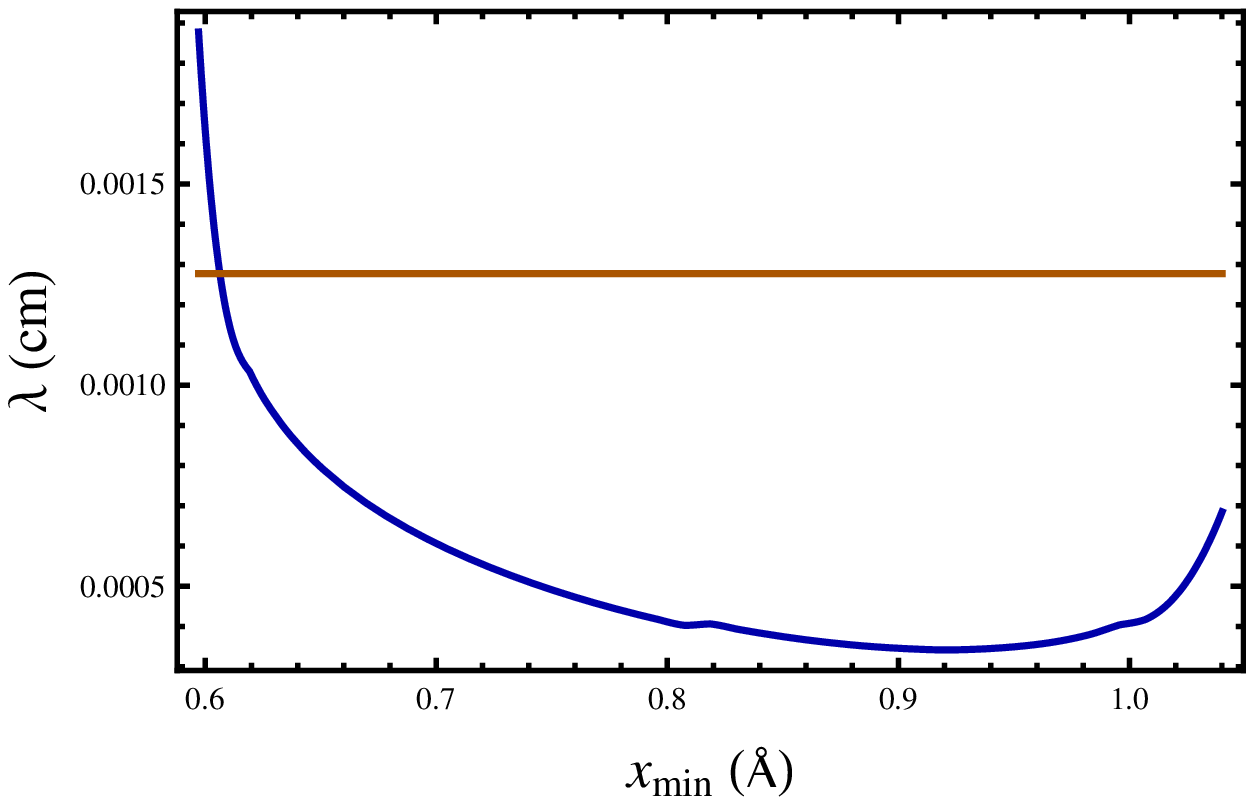}}
 \caption{\label{F3} Channeling oscillation length for 20.35GeV electrons, $l_{cr}=0.11 mm$, $\theta_b=1600\mu rad$, (110) planes.}
 \end{figure}

For the negative particles, the interplanar potential $U(x)$ becomes inverted, resulting in the atomic plane being in the channel center. Thus, the dechanneling zone includes all the amplitudes of oscillations. Depending on the amplitude, the channeling oscillation length can differ several times as is shown in Fig. \ref{F3}. In addition, electrons cross crystal planes in the middle of the channel when their angle $\theta$ is maximal in magnitude \cite{Tikh2}. Positrons, on the opposite, approach the planes at the minimal angle values. Consequently the transverse energy change $\Delta\varepsilon_\bot$ for electrons is proportional to the scattering angle $\Delta\theta$ while for positrons to its square $\Delta\theta^2$ \cite{Tikh2}:
\begin{eqnarray}
\label{70}
\Delta\varepsilon_\bot=pv\theta\Delta\theta+pv\frac{\Delta\theta^2}{2}\rightarrow
\begin{cases}
pv\theta\Delta\theta,&\text{if $\theta\rightarrow\theta_{max}$;}\\
pv\frac{\Delta\theta^2}{2},&\text{if $\theta\rightarrow0$.}
\end{cases}
\end{eqnarray}

Thereby, the amplitude of electron oscillations due to scattering increases stronger for electrons than for positrons. Thus, any phase correlation will quickly disappear and the planar channeling oscillations for electrons will not be observable in the angular distribution.

It is also important to explain why the pattern of peaks is a sequence of a high peak followed by a lower one (see Fig. \ref{F0}).  This is explained by an asymmetry of the potential well displayed in Fig. \ref{F2}. In particular, the dechanneling zone close to the left side of the potential is wider than the zone near the opposite reflection point. Indeed, if one takes the dechanneling zone width to be equal to the thermal vibration amplitude (0.075${\AA}$ for (110) silicon crystal planes), one obtains the corresponding potential energy difference $\Delta U \sim$ 2 eV. Its value as well as the dechanneling zone width does not considerably change for different crystal bending and beam energies. In contrast, the width of the zone near the reflection point $x_{ref}$ (see Fig. \ref{F2}) strongly depends on the crystal bending:
\begin{equation}
\label{8}
\Delta x \approx \Delta U /U_{eff}'(x_{ref}).
\end{equation}
Through the use of the numerical parameters of the potential in Fig. \ref{F2}d, one obtains $\Delta x=0.042 {\AA}$, which is almost two times less than the thermal vibration amplitude. This ratio explains the alternation of high and low peaks.

Planar channeling oscillations can also be observed for (111) crystal planes. The main contribution here is due to the wider channel because the dechanneling zone introduced above is considerably narrower for the small channel as is shown in Fig. \ref{F2}f. In addition, the channeling oscillation length varies more strongly in the latter case, causing smearing of phase correlations. The oscillation length value is also smaller for the small channel than for the larger one. This causes reduction of the distance between the peaks, resulting in complicated detection. Thereby, only the planar channeling oscillations in the wider channels will be practically observable.

The appearance of dechanneling peaks is also possible for volume-captured particles when the volume reflection orientation is set. In this case, the lower peaks will completely disappear because volume capture occurs only near the left potential maximum (see Fig. \ref{F2}). For this reason, the difference between the peaks corresponds to one channeling oscillation length for the volume reflection orientation.

The dechanneling peaks may be observed if the inter-peak half angular distance $\Delta \varphi _{ch}$ exceeds the doubled incoherent scattering angle $\theta _{sc}$ \cite{Sytov}:
\begin{equation}
\label{2}
\frac{\Delta \varphi _{ch}}{2 \theta _{sc}}=\frac{\lambda  \theta _b}{4 l_{cr}}\frac{pv}{13.6 MeV \sqrt{l_{cr}/X_r} \left(0.038 \ln \left(l_{cr}/X_r\right)+1\right)}>1,
\end{equation}
where $l_{cr}$ and $\theta_{b}$ are the crystal length and bending angle, respectively, $X_{r}$ is the radiation length equal to 9.36 cm for silicon. The Coulomb scattering angle was estimated according to \cite{PDG}. For the volume reflection orientation this condition will be twice softer because half of the peaks will not appear as mentioned above. Thus, volume-reflection orientation provides a still better conditions than that for channeling.

Another important condition is the crystal bending radius $R$ to be larger than the critical one, $R_{cr}$ \cite{Biryukov}:
\begin{equation}
\label{3}
\frac{R}{R_{cr}}>1.
\end{equation}
As mentioned above, this is the condition for the channeling to occur in a bent crystal \cite{Tsyganov, Biryukov}.

The third important condition is the angular divergence of the incident beam should not be greater than half of the critical channeling angle $\theta_L$.
\begin{equation}
\label{31}
\theta_{in \textcolor[rgb]{1.00,1.00,1.00}{p}r.m.s.} < \theta_L/2.
\end{equation}
Indeed, the angular divergence, approaching the critical angle, results in a considerable oscillation phase shift. Such trajectories are, of course, uncorrelated. This concerns both channeling and volume reflection orientations.

Some sort of scaling of the channeling oscillation picture with energy can readily be introduced. Such a scaling can be assured by the conservation of both the peak number:
\begin{equation}
\label{4}
n_{peaks}=\frac{2 l_{cr}}{\lambda}=Const;
\end{equation}
and of the ratio of the inter-peak interval to the Coulomb scattering angle:
\begin{equation}
\label{5}
\frac{\Delta \varphi _{ch}}{2 \theta _{sc}}=Const.
\end{equation}
By substituting Eq. (\ref{1}) into (\ref{4}) one obtains:
\begin{equation}
\label{6}
l_{cr}\sim\sqrt{pv}.
\end{equation}
Substituting further Eq. (\ref{2}) into (\ref{5}), using (\ref{6}) and neglecting the logarithmic factor one obtains that:
\begin{equation}
\label{7}
\theta_{b}\sim1/(pv)^{3/4}.
\end{equation}
Finally the bending radius scaling can be simply obtained from Eqs. (\ref{6}) and (\ref{7}):
\begin{equation}
\label{71}
R\sim (pv)^{5/4}.
\end{equation}

\subsection{Quasichanneling oscillations}

Planar channeling oscillations in backscattering experiments at low energy were observed \cite{Abel1,Abel2,Abel3,Kauf,PCO1,PCO2,Berec} while they have not been observed yet at higher energy. However, there is another kind of oscillations, which we predict in this paper that has not still observed under neither regimes. This kind of oscillations manifests itself as the peaks in the angular distribution which are close to the channeling peak (Fig. \ref{F0}). Hereinafter, such new kind of oscillations will be called \textit{planar quasichanneling oscillations}.

Such oscillations have a different nature than planar channeling oscillations because the distance between them is smaller than the lowest possible half channeling length. In addition, the location of the peaks is almost the same for particles with different charge signs. As we will show below, this indicates the involvement of over-barrier particles.

\begin{figure}
\resizebox{77mm}{!}{\includegraphics{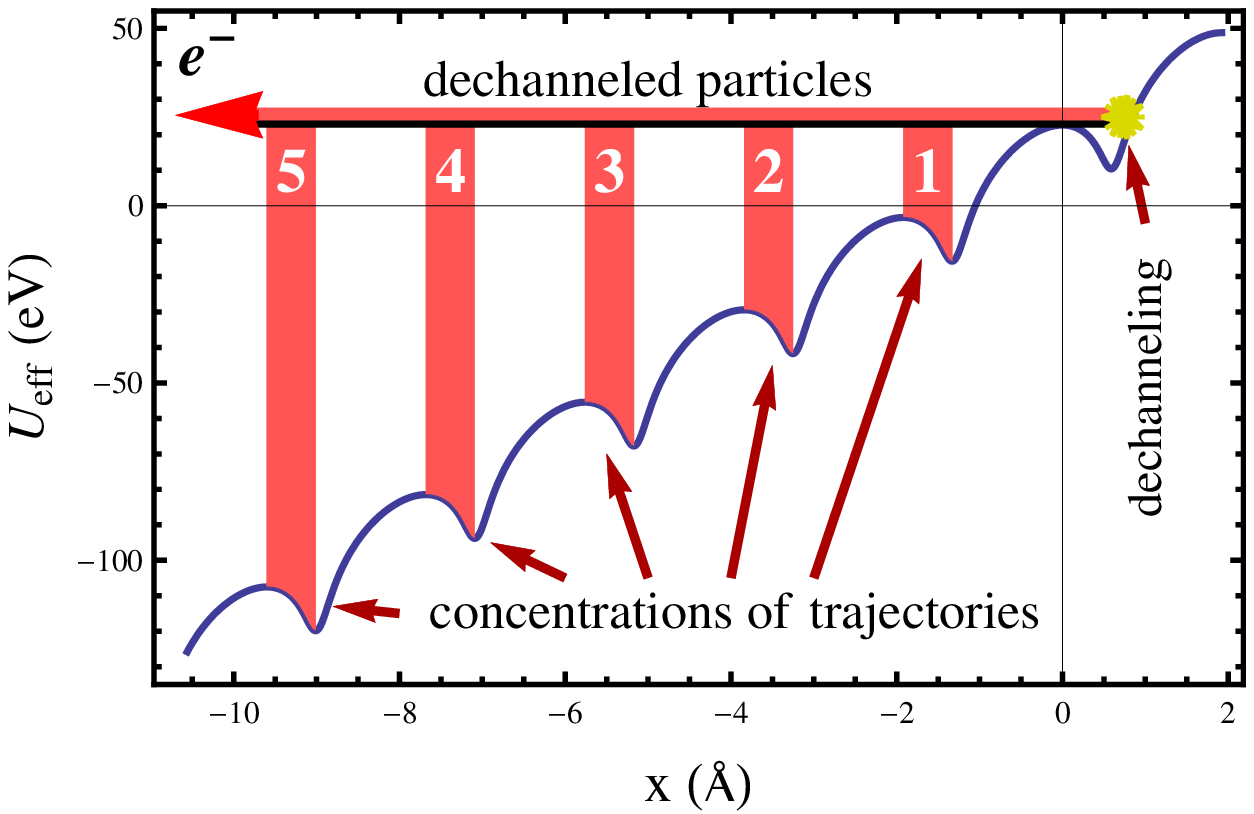}}
\resizebox{77mm}{!}{\includegraphics{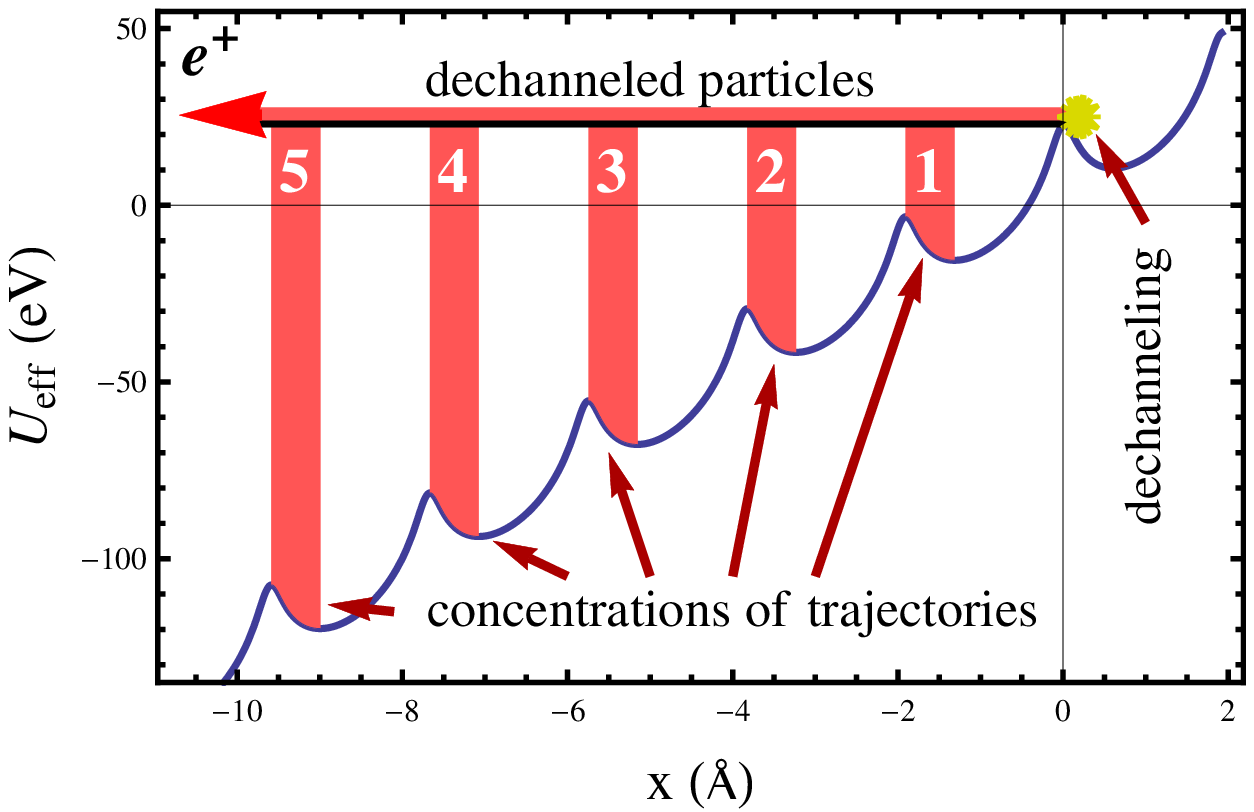}}
\resizebox{77mm}{!}{\includegraphics{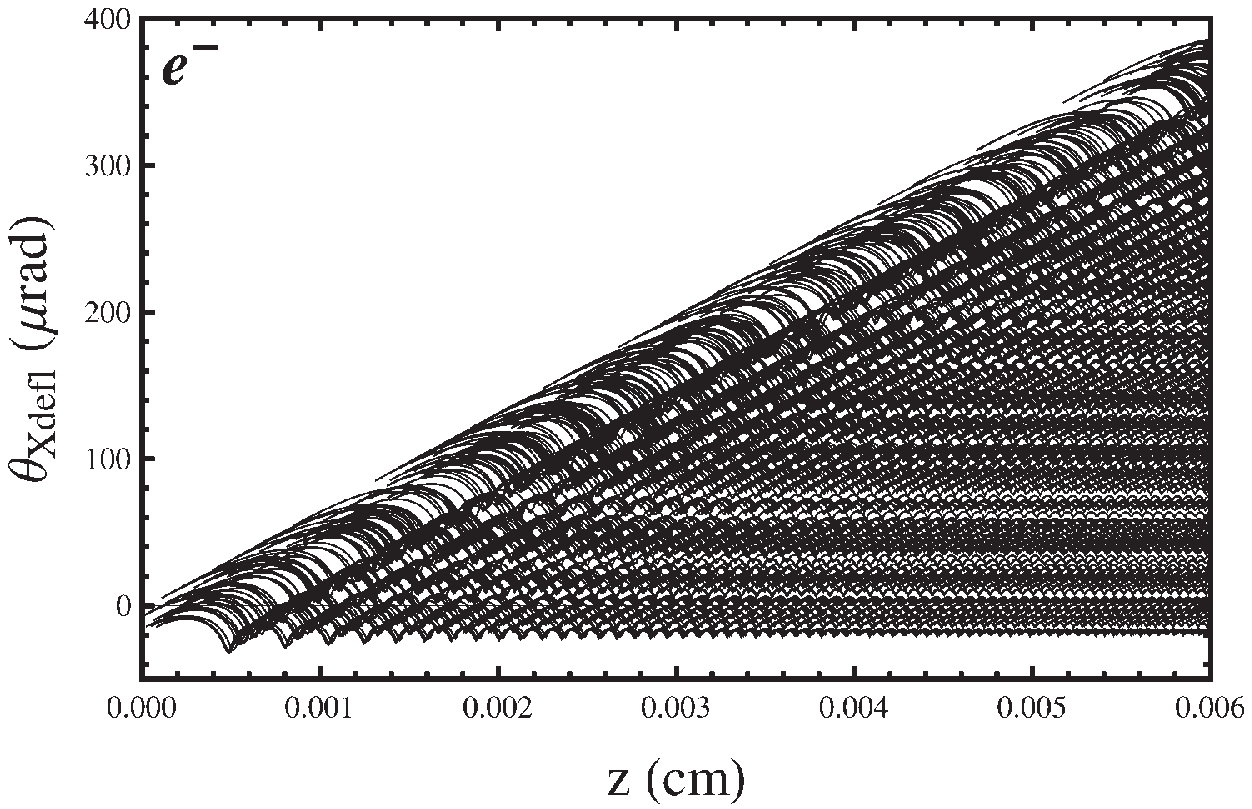}}
\resizebox{77mm}{!}{\includegraphics{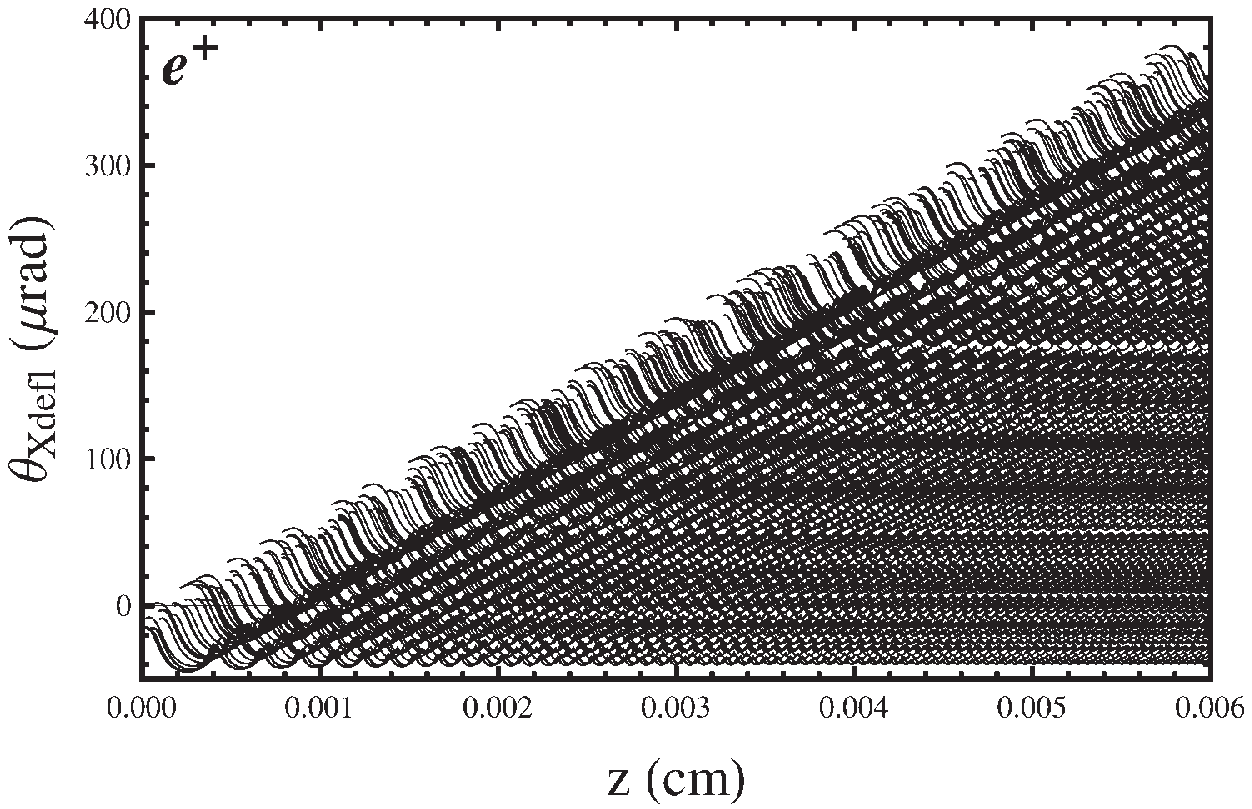}}
\resizebox{77mm}{!}{\includegraphics{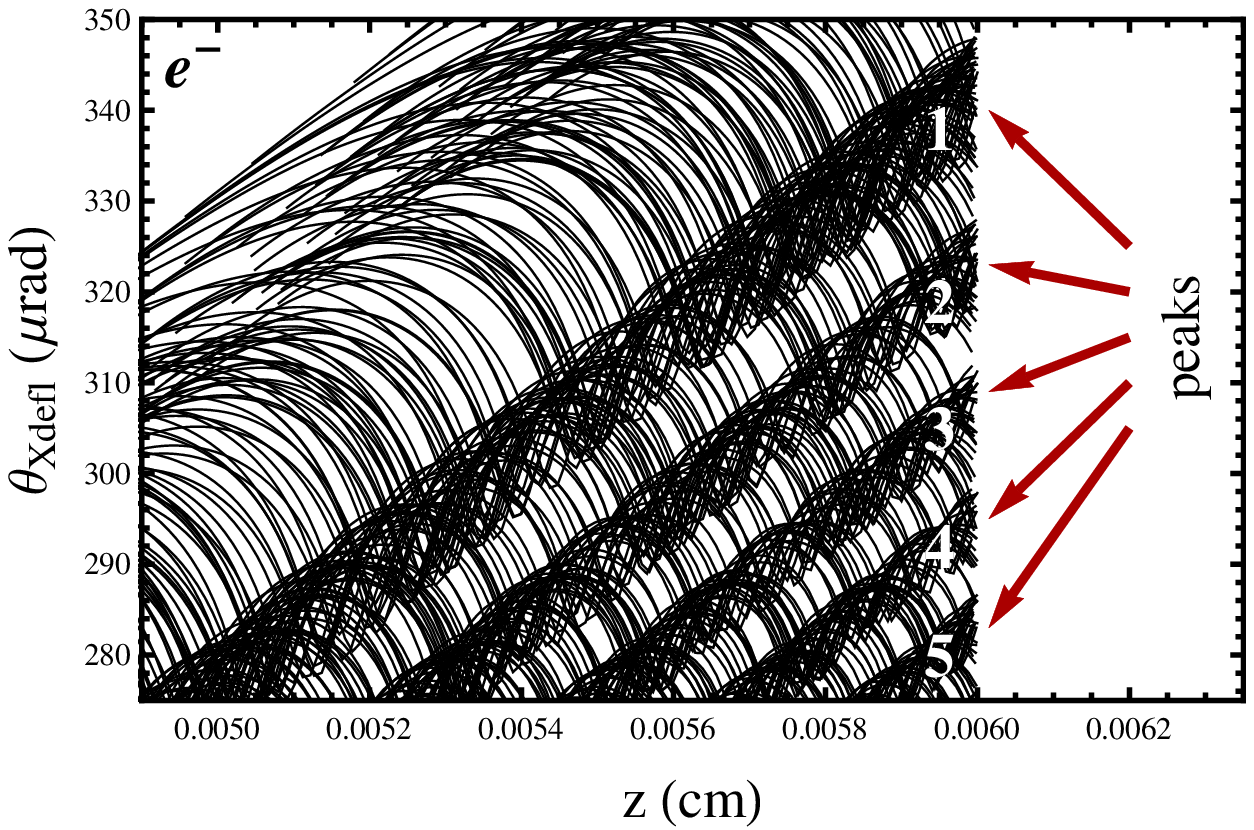}}
\resizebox{77mm}{!}{\includegraphics{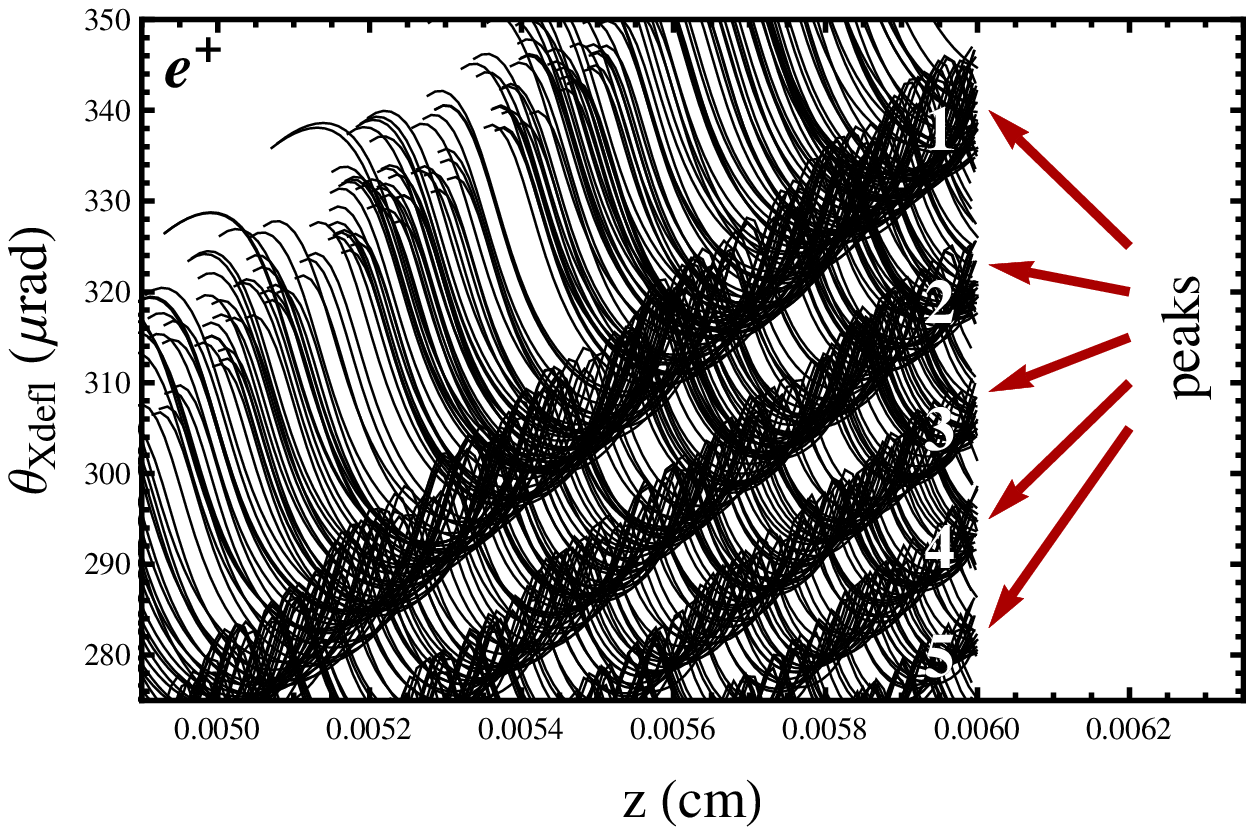}}
 \caption{\label{F6} The effective potential $U_{eff}$ (top) and the ideal over-barrier trajectories (middle and bottom) (angle (\ref{12}) \textit{vs} longitudinal coordinate) of 20.35 GeV electrons (left column) and for positrons (right column) without scattering. The transverse starting point of trajectories is fixed at $x=0$, the longitudinal one varies randomly. The transverse energy varies in the range of several eV above the potential barrier at $x=0$. The crystal parameters are: $l_{cr}=60 \mu rad$, $\theta_b=400\mu rad$, (110) planes. The longitudinal coordinates in bottom figures are close to the crystal end.}
 \end{figure}

A qualitative explanation can be obtained from the analysis of the over-barrier trajectories, shown in Fig. \ref{F6}, highlighting the dependence of the particle deflection angles in the laboratory reference system on $z$:
\begin{eqnarray}
\begin{array}{l}
\theta_{Xdefl}=\frac{z}{R}-\sqrt{\frac{2(\varepsilon_\bot-U_{eff}(x(z,\varepsilon_\bot)))}{pv}},
\end{array}
\label{12}
\end{eqnarray}
where $\varepsilon_\bot$ is the initial transverse particle energy, $x$ and $z$ the particle transverse and longitudinal coordinates respectively. At $z=l_{cr}$, the first term in Eq. (\ref{12}) becomes $z/R=\theta_b$ and $\theta_{Xdefl}$ is the observed particle deflection angle as in Fig. \ref{F0}.

Let us consider the ideal trajectories without any incoherent scattering or energy losses. Also, only dechanneled particles are considered because usually most of the particles initially not captured under channeling state will not achieve the angles close to the channeling direction. Let us also fix the starting point of the over-barrier trajectories in a point above the potential barrier, neighboring the dechanneling point (in Fig. 6 the point is indicated as $x=0$). We will vary randomly only the longitudinal starting coordinate. Fig. \ref{F6} demonstrates that the trajectories tend to group together at certain phases into parallel lines separated by one over-barrier oscillation. 
Such concentration will generate a new series of peaks in the angular distribution at the crystal exit.

All the lines formed are parallel to the line representing the angle of bending of a crystal plane:
\begin{eqnarray}
\begin{array}{l}
\theta_{Xdefl}=z/R.
\end{array}
\label{125}
\end{eqnarray}
The main reason for their appearance, is correlation of different over-barrier trajectories even in the first potential well (see Fig. \ref{F6}). These trajectories have almost the same oscillation lengths with the only exception of a small region near the closest barrier to the point of dechanneling. Therefore, all the over-barrier trajectories differ only by a starting longitudinal coordinate which varies along a bent crystal plane parallel to the line (\ref{125}) according to Eq. (\ref{12}).

The oscillation length of different trajectories is almost the same by the reason that the main contribution to the quasichanneling peaks is due to dechanneling process. The transverse energy of dechanneled particles is limited by Coulomb scattering and, therefore, can exceed the closest potential barrier, at least by several eV. The relative change of the n-th over-barrier oscillation length can be calculated by formula:
\begin{eqnarray}
\begin{array}{l}
\frac{\Delta\lambda_n}{\lambda_n}=\frac{\Delta\varepsilon_\bot}{2}\frac{\int_{(n-1)d_0}^{nd_0} \frac{dx}{(\varepsilon_\bot-U_{eff}(x))^{3/2}}}{\int_{(n-1)d_0}^{nd_0} \frac{dx}{\sqrt{\varepsilon_\bot-U_{eff}(x)}}}.
\end{array}
\label{127}
\end{eqnarray}
As a rule, this ratio does not exceed $\sim$ 10\% for dechanneled particles. However, this also relates to the particles, initially not captured under the channeling mode but achieved the deflection angle close to the channeling direction. Thereby, they must provide the peaks of quasichanneling oscillations to the same locations as the dechanneled ones.

The location of parallel lines can be found by the condition that the tangent lines $d\theta_{Xdefl}/dz$ to the trajectories are parallel to the line (\ref{125}). This condition transforms to:
\begin{eqnarray}
\begin{array}{l}
\frac{dU_{eff}}{dx}=0,
\end{array}
\label{128}
\end{eqnarray}
which implies the locations of local minima and maxima of the potential $U_{eff}$ (see Fig. \ref{F6}). Therefore, the trajectories group between the minima and maxima as shown in Fig. \ref{F6} because $d\theta_{Xdefl}/dz\simeq z/R$.

By application of the potential values $U_{eff}$ for minima and maxima and using Eq. (\ref{12}), one obtains the equations of two parallel lines which are the boundaries of trajectory concentrations:
\begin{eqnarray}
\begin{array}{l}
\theta_{Xdefl}=z/R-\sqrt{\frac{2V_0n}{pv}};\\
\theta_{Xdefl}=z/R-\sqrt{\frac{2(V_0n+\Delta V)}{pv}},
\end{array}
\label{13}
\end{eqnarray}
for the potential maxima and minima respectively. $\Delta V$ is the potential energy difference between the neighboring local maximum and minimum while $V_0$ is the difference between two neighboring maxima of the potential \cite{Biryukov}:
\begin{eqnarray}
\begin{array}{l}
V_0=pvd_{0}/R.
\end{array}
\label{16}
\end{eqnarray}

By substituting Eq. (\ref{16}) in (\ref{13}) and taking into account $z=l_{cr}$ at the crystal exit one finally obtains the location of the bounds containing the peaks of quasichanneling oscillations in the deflection angle distribution:
\begin{eqnarray}
\begin{array}{l}
\theta_{Xdefl}=\theta_b-\sqrt{\frac{2d_0n}{R}};\\
\theta_{Xdefl}=\theta_b-\sqrt{\frac{2d_0n}{R}+\frac{2\Delta V}{pv}},
\end{array}
\label{13}
\end{eqnarray}
Negative particles tend to be closer to the first angle while the positive to the second one, where the derivative $d\theta_{Xdefl}/dz$ is smoother. The angular difference between neighboring peaks $\Delta\varphi_{qch}$ can be found from these equations as:
\begin{eqnarray}
\begin{array}{l}
\Delta\varphi_{qch}=\sqrt{\frac{2d_{0}}{R}+(\theta_{b}-\theta_{Xdefl})^2}-(\theta_{b}-\theta_{Xdefl}).
\end{array}
\label{17}
\end{eqnarray}
For large $n$, this equation reduces into:
\begin{eqnarray}
\begin{array}{l}
\Delta\varphi_{qch}\approx\frac{d_{0}}{R(\theta_{b}-\theta_{Xdefl})}.
\end{array}
\label{18}
\end{eqnarray}
It is important to stress that such formula does not depend on the particle energy but only on crystal characteristics, such as interplanar distance and bending radius.

Being an over-barrier effect, quasichanneling oscillations can be experimentally observed for any angular divergence less, of course, than the crystal bending angle. The main constraint here is the limited statistics of the over-barrier particles in the angular distribution, which depends in turn on channeling efficiency. Therefore, short crystals are preferred to provide the highest efficiency.

In order to find the extremal conditions where the observation of quasichanneling oscillations is still possible, one can estimate only the first oscillation forming the closest peak to the channeling one. For the initial angle $\theta_{Xdefl}$ in (\ref{18}) one should take the left boundary of the channeling peak to be $\theta_{b}-\theta_{L}$. In this case one obtains the highest possible angular difference between the channeling peak and the peak of a quasichanneling oscillation:

\begin{equation}
\label{19}
\frac{\Delta\varphi_{qch}}{2 \theta _{sc2}}=\frac{d_{0}}{2R\theta_{L}}\frac{pv}{13.6 MeV \sqrt{\frac{\lambda_1}{X_r}} \left(0.038 \ln \left(\frac{\lambda_1}{X_r}\right)+1\right)}>1.
\end{equation}
For an estimate, $\lambda_1$ can be roughly estimated to be equal to half of the channeling oscillation length (\ref{1}), i.e. $\lambda_1\sim\lambda/2$.

In order to observe quasichanneling oscillations, one should also satisfy the condition of the bending radius to be larger than the critical one (\ref{3}).

Since Eq. (\ref{19}) for quasichanneling oscillations depends on energy like Eq. (\ref{7}) for channeling oscillations, they scale on energy in the same way.

The ratio of maximal interpeak distance of quasichanneling oscillations to that of channeling oscillations can be estimated by using (\ref{1}), (\ref{2}), (\ref{31}) and (\ref{19}), resulting in:

\begin{equation}
\frac{\Delta \varphi _{qch}}{\Delta \varphi _{ch}}<\frac{2}{\pi}.
\label{21}
\end{equation}
For volume reflection, the analogous ratio will be two times lower. Thus, the width between the peaks for channeling oscillations is considerably higher than for the quasichanneling ones. However, the different role of multiple scattering (compare (\ref{2}) and (\ref{19})) and angular divergence makes the conditions for observation of quasichanneling generally more preferable.

\section{Simulation results}

For a deeper understanding of both channeling and quasichanneling oscillations we performed a numerical simulation.

\begin{figure}
\resizebox{77mm}{!}{\includegraphics{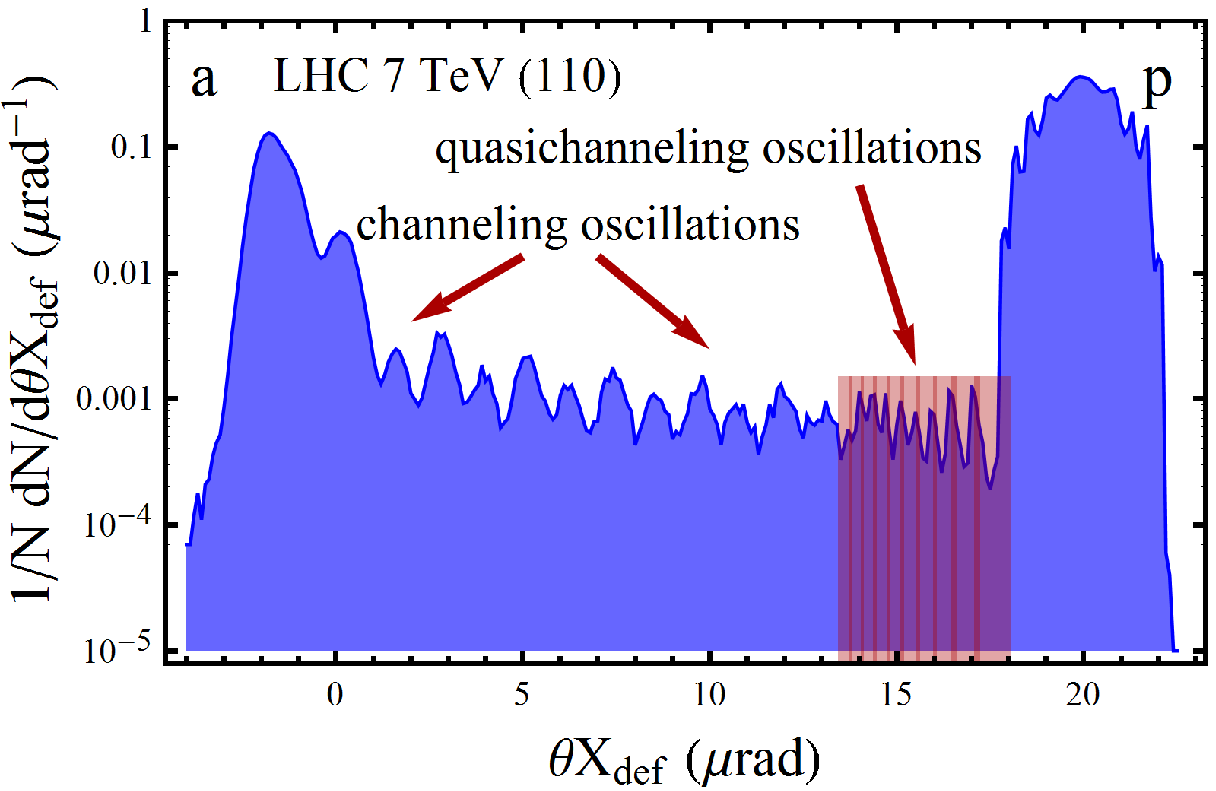}}
\resizebox{77mm}{!}{\includegraphics{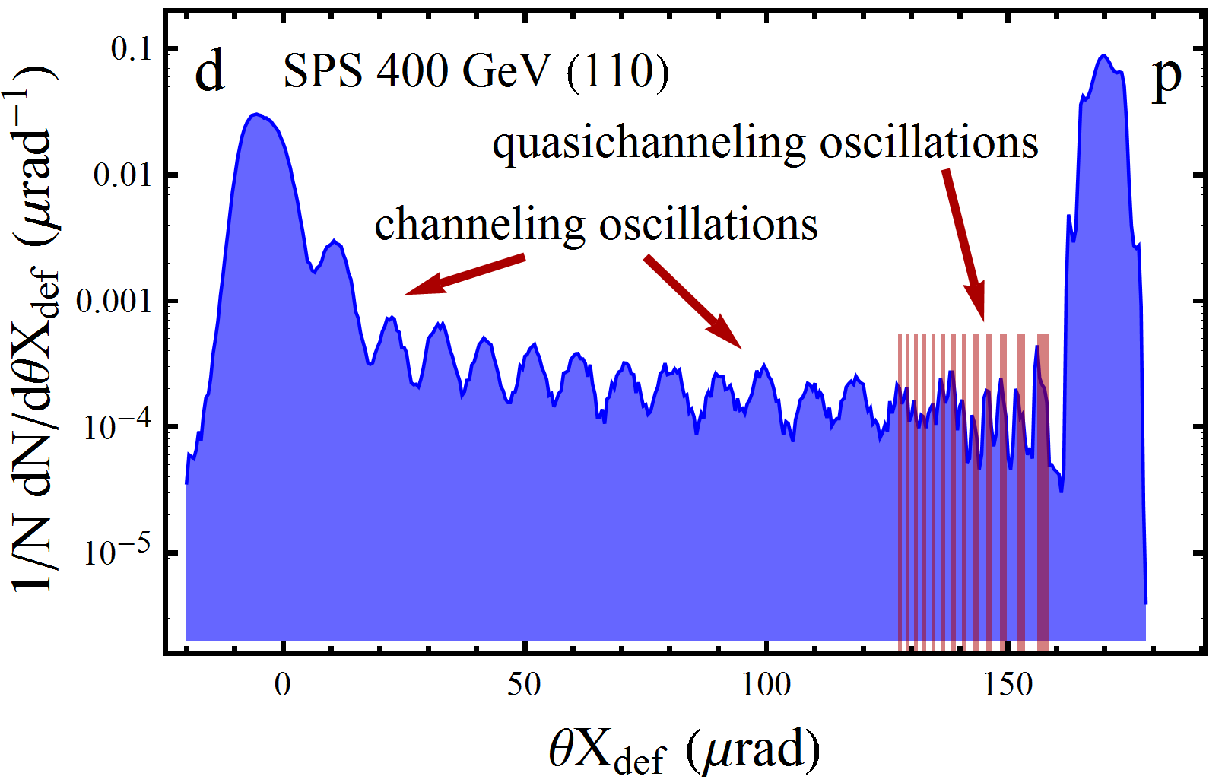}}
\resizebox{77mm}{!}{\includegraphics{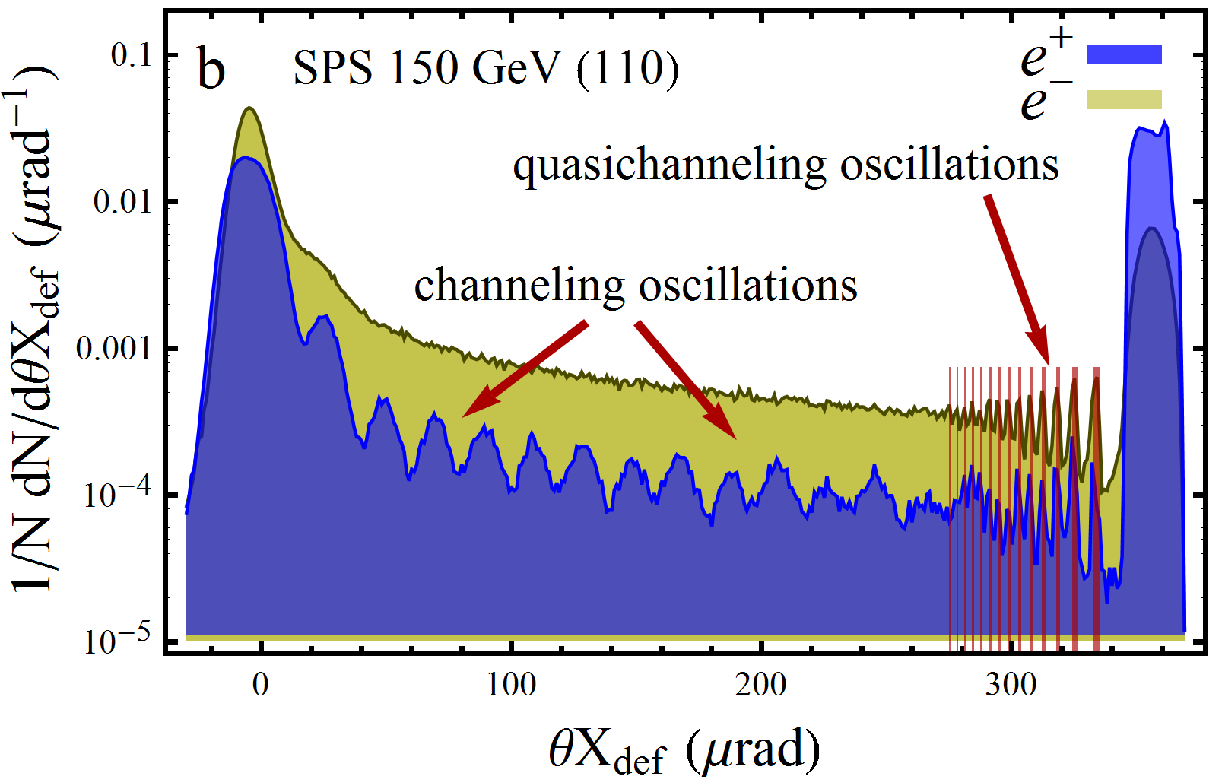}}
\resizebox{77mm}{!}{\includegraphics{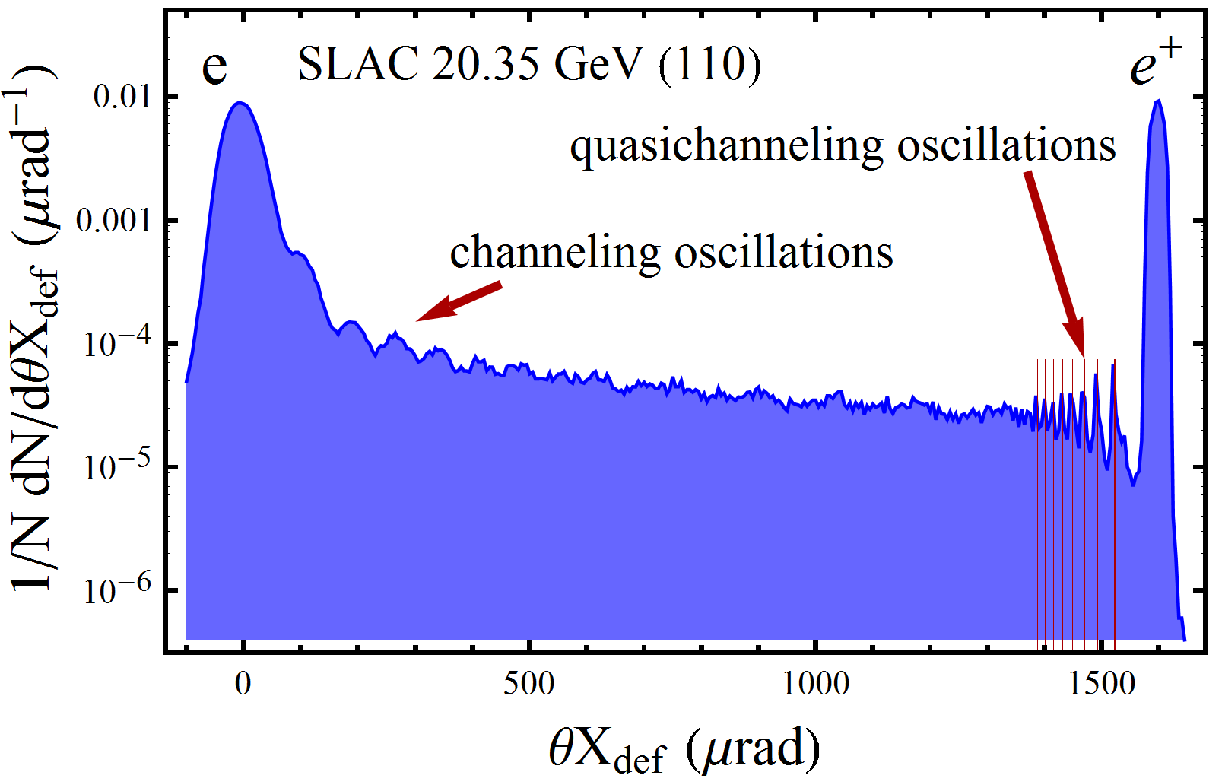}}
\resizebox{77mm}{!}{\includegraphics{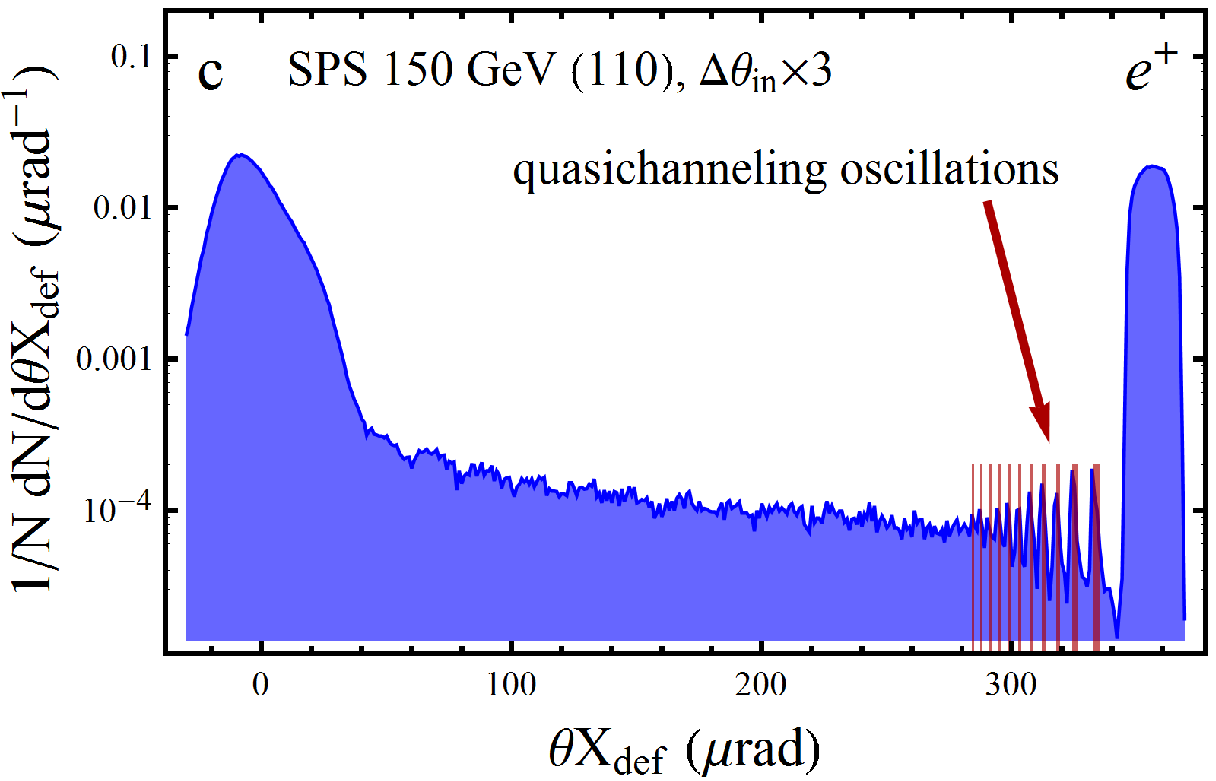}}
\resizebox{77mm}{!}{\includegraphics{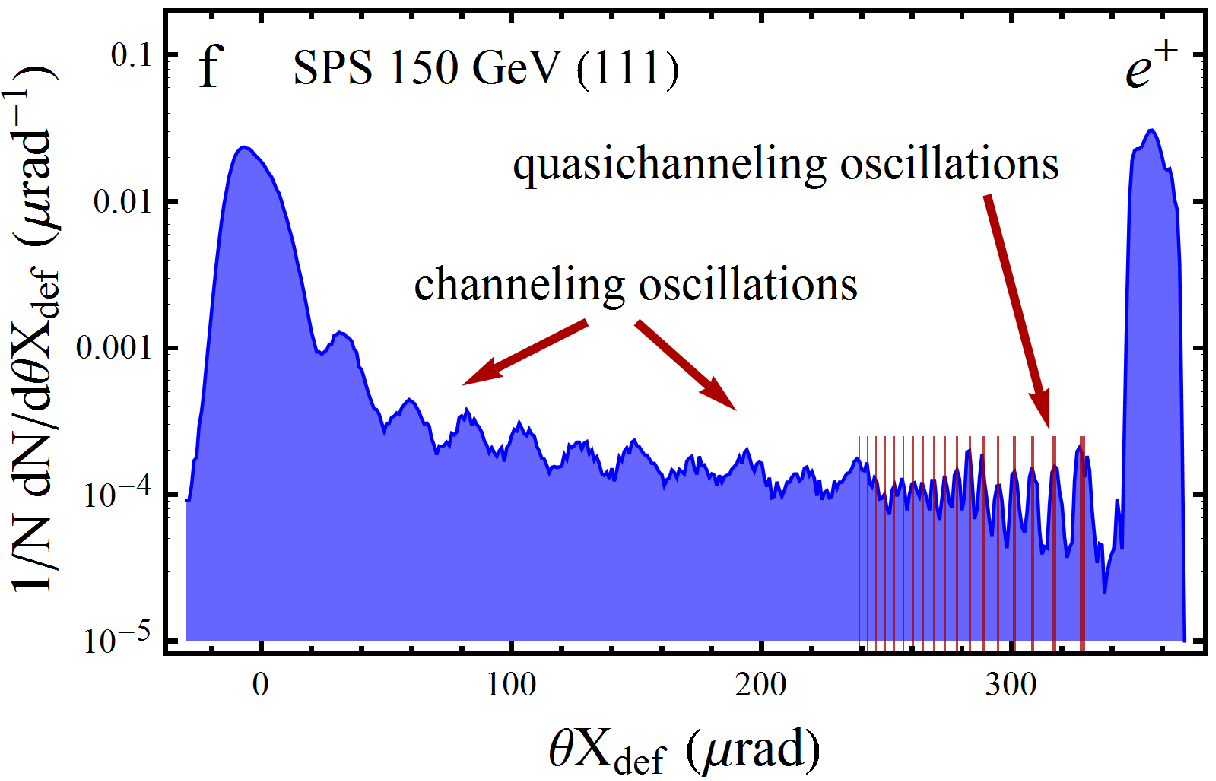}}
 \caption{\label{F1} The angular distributions of particles after interaction with the silicon crystal at the channeling orientation. Vertical lines indicate the zone of quasichanneling oscillations manifestation calculated by Eq. (\ref{13}). The simulation layouts were as follows: (a) LHC, 7 TeV protons, r.m.s. beam angular divergence $\theta_{in}=0.5\mu rad$, $l_{cr}=2 mm$, $\theta_b=20\mu rad$, (110) planes; (b) SPS, 150 GeV positrons and electrons, $\theta_{in}=3.5\mu rad$, $l_{cr}=0.29 mm$, $\theta_b=357\mu rad$, (110) planes; (c) the same as the previous except the angular divergence $\theta_{in}=10.5\mu rad$; (d) SPS, 400 GeV protons, $\theta_{in}=2\mu rad$, $l_{cr}=0.48 mm$, $\theta_b=170\mu rad$, (110) planes; (e) SLAC, 20.35 GeV positrons, $\theta_{in}=10\mu rad$, $l_{cr}=0.11 mm$, $\theta_b=1600\mu rad$, (110) planes; (f) the same as (b) for (111) planes.}
 \end{figure}

\begin{figure}
\resizebox{77mm}{!}{\includegraphics{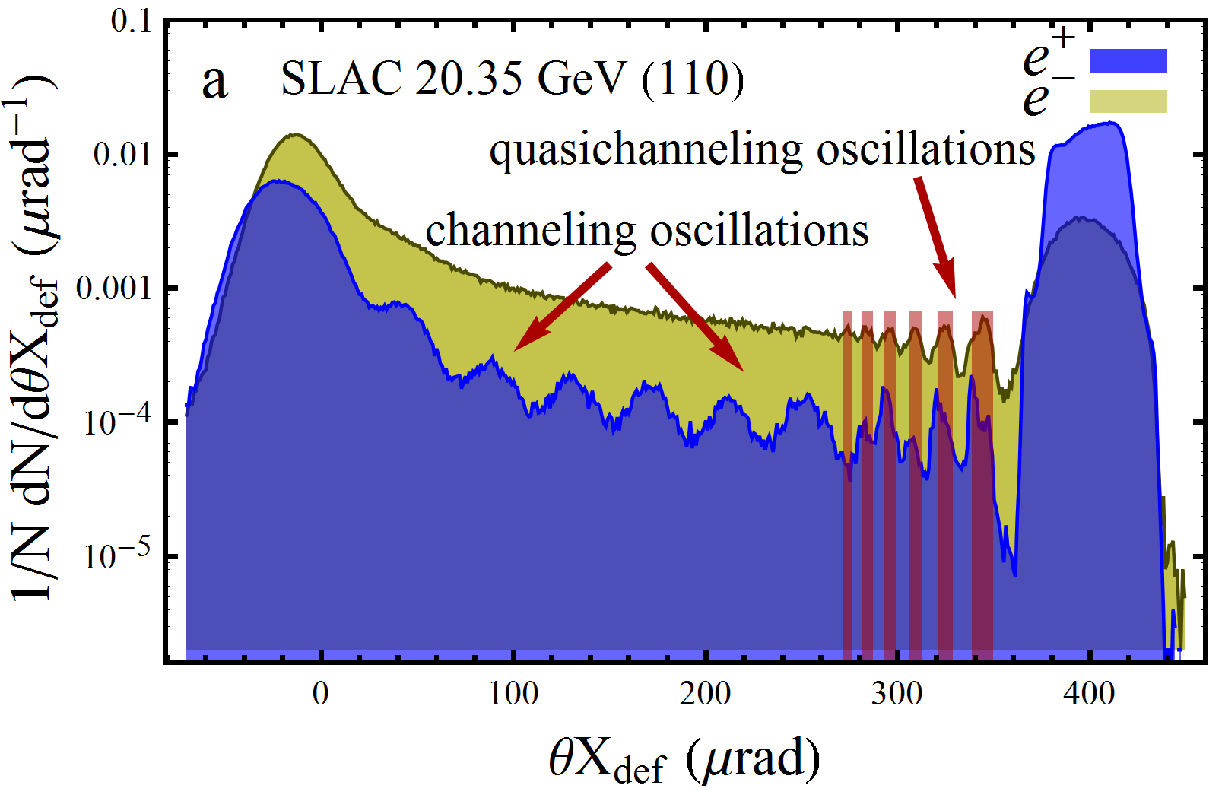}}
\resizebox{77mm}{!}{\includegraphics{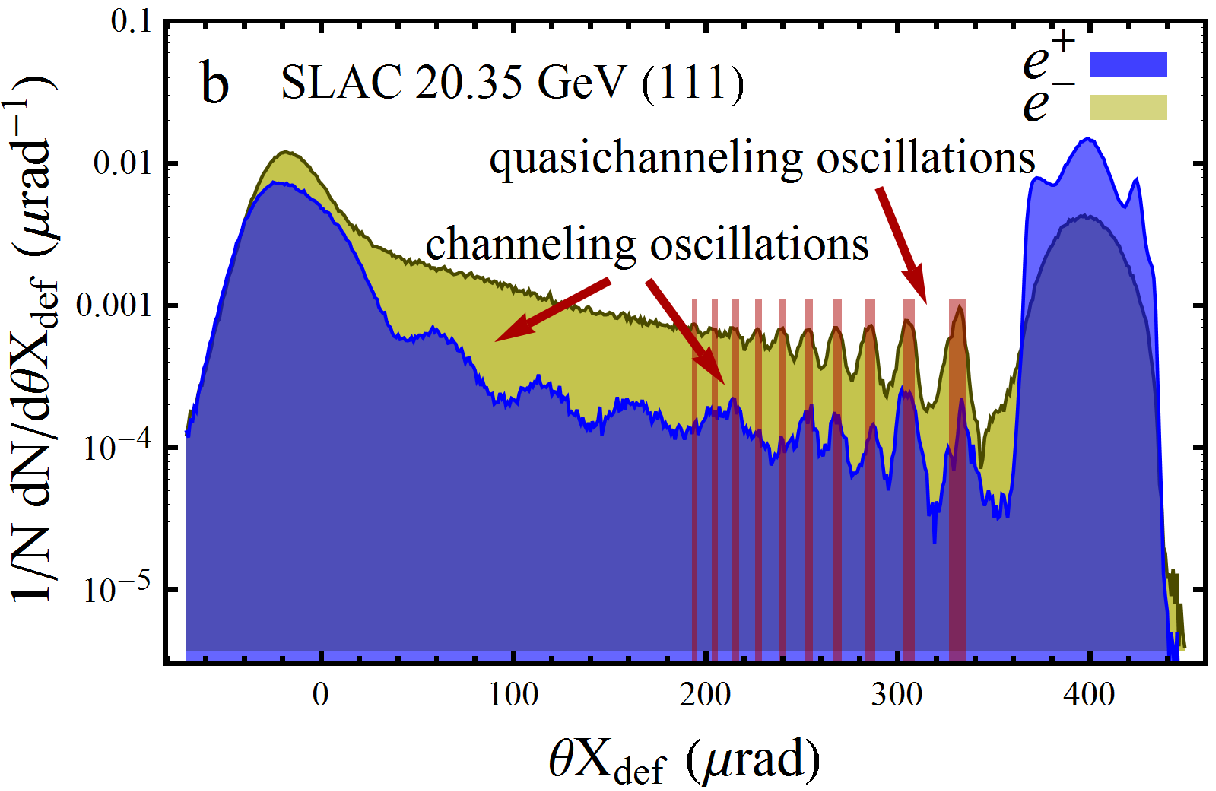}}
 \caption{\label{F11}(a) SLAC, 20.35 GeV positrons and electrons, $\theta_{in}=10\mu rad$, $l_{cr}=60 \mu m$, $\theta_b=400\mu rad$, (110); (b) the same as (g) for (111) planes.}
 \end{figure}

 \begin{figure}
\resizebox{77mm}{!}{\includegraphics{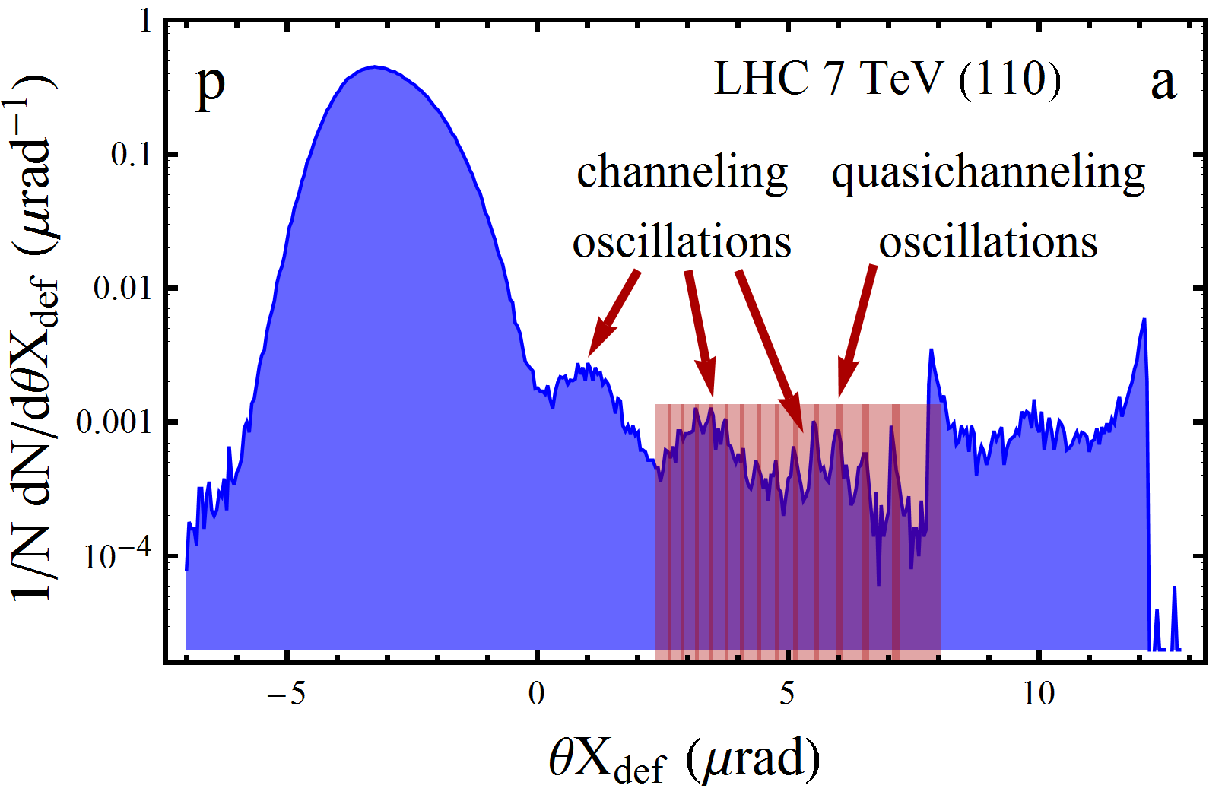}}
\resizebox{77mm}{!}{\includegraphics{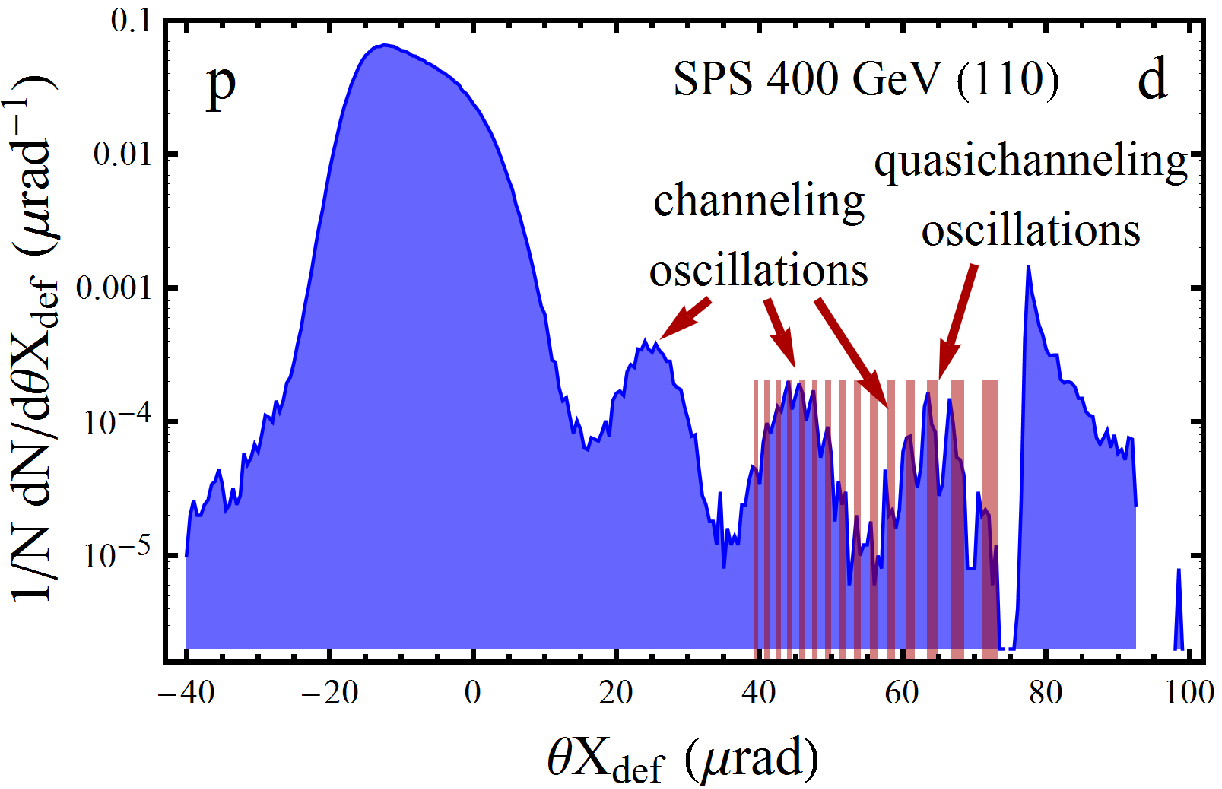}}
\resizebox{77mm}{!}{\includegraphics{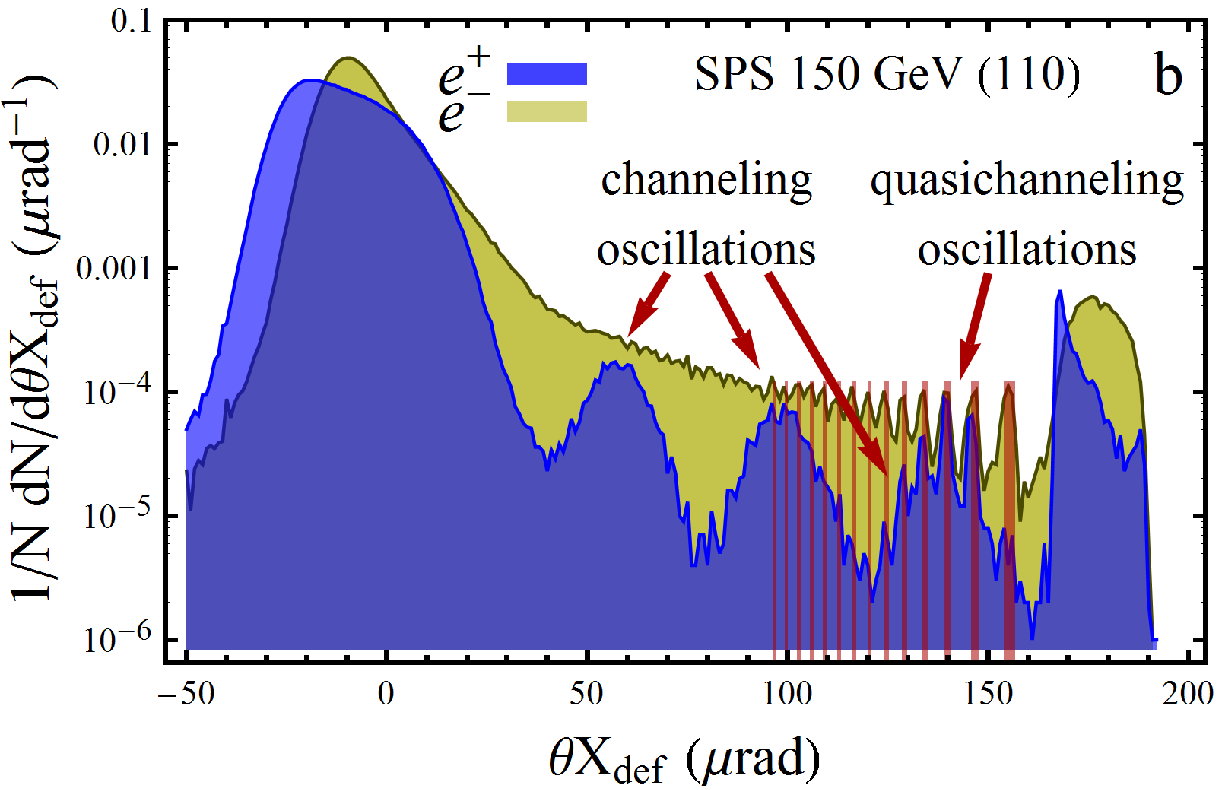}}
\resizebox{77mm}{!}{\includegraphics{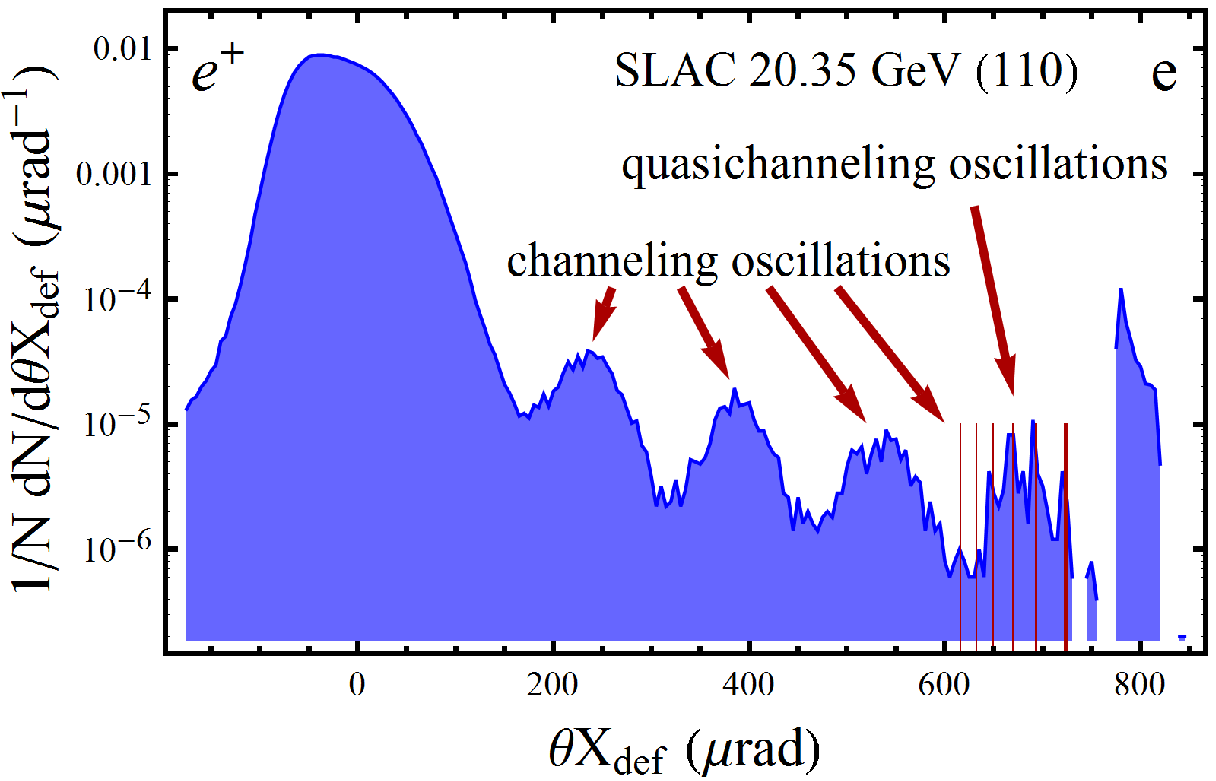}}
\resizebox{77mm}{!}{\includegraphics{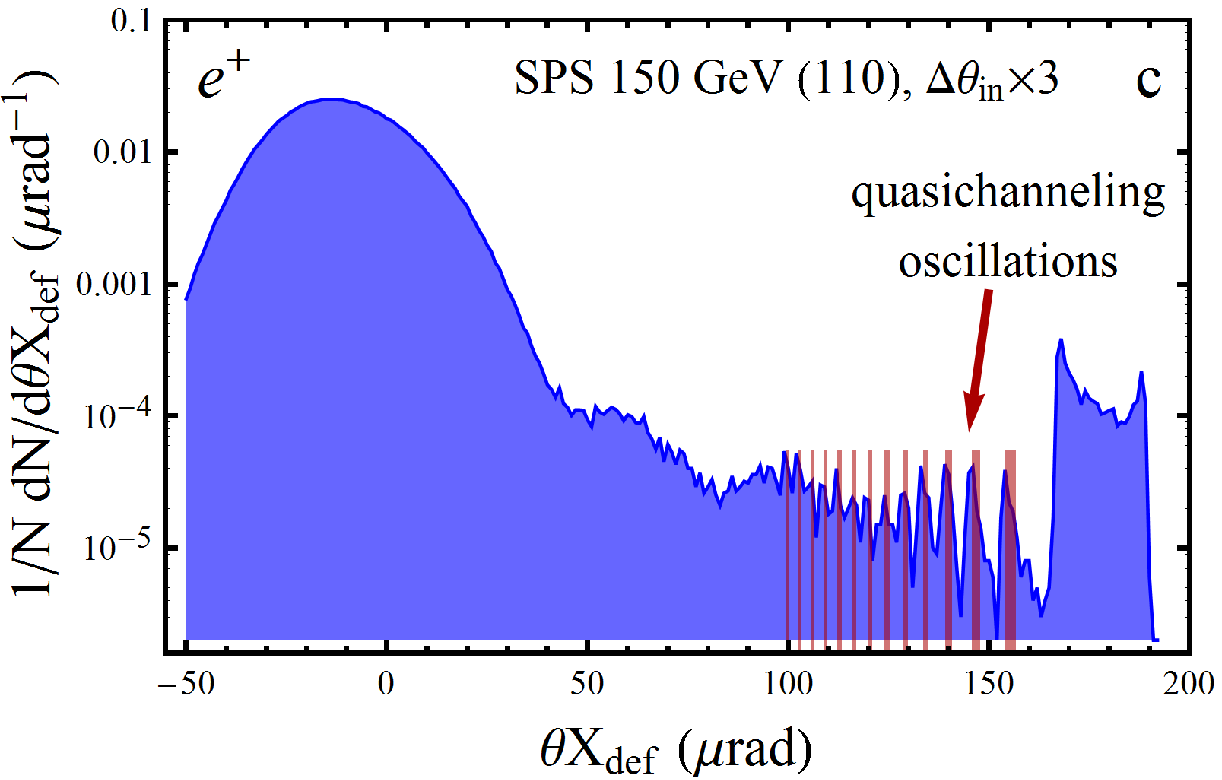}}
\resizebox{77mm}{!}{\includegraphics{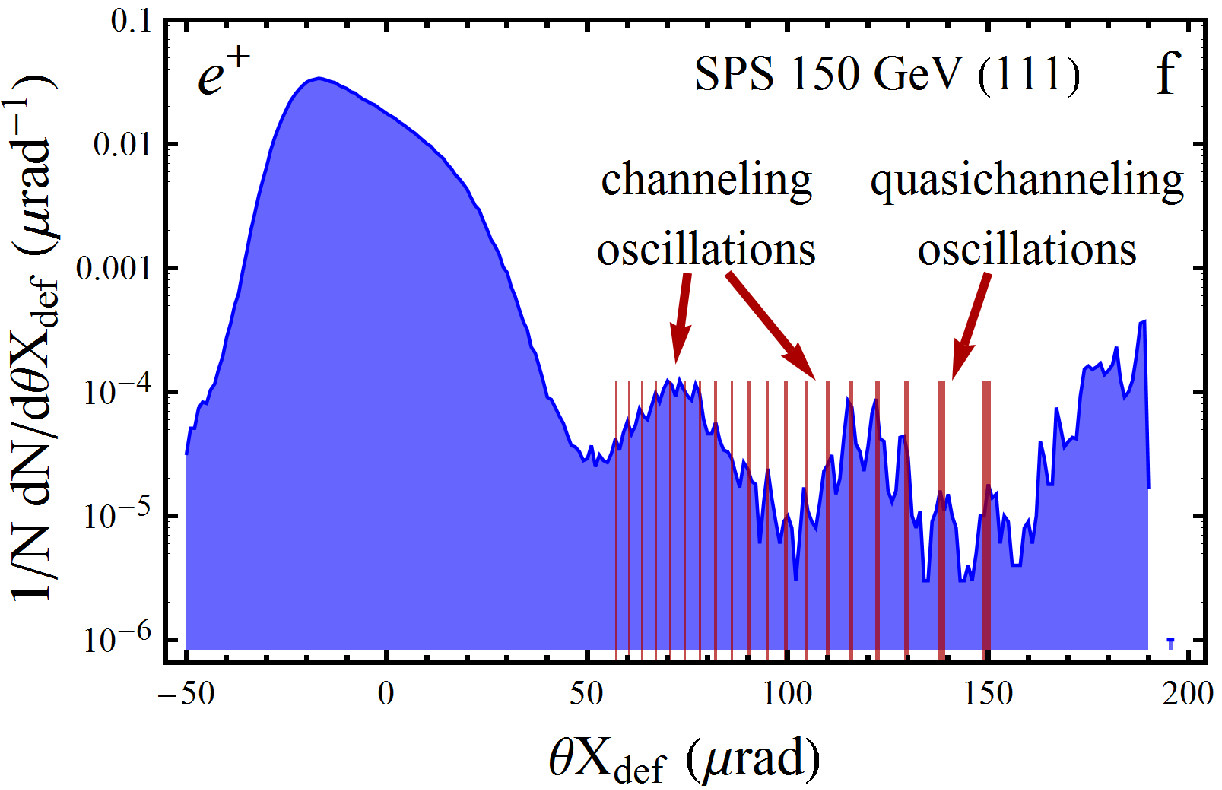}}
 \caption{\label{F4} The angular distributions of particles after interaction with the silicon crystal at the volume reflection orientation for the crystal tilt $-\theta_b/2$. Vertical lines indicate the zone of quasichanneling oscillations manifestation calculated by Eqs. (\ref{13}). The simulation layouts were as follows: (a) LHC, 7 TeV protons, r.m.s. beam angular divergence $\theta_{in}=0.5\mu rad$, $l_{cr}=2 mm$, $\theta_b=20\mu rad$, (110) planes; (b) SPS, 150 GeV positrons and electrons, $\theta_{in}=3.5\mu rad$, $l_{cr}=0.29 mm$, $\theta_b=357\mu rad$, (110) planes; (c) the same as (b) for the angular divergence $\theta_{in}=10.5\mu rad$ for positrons; (d) SPS, 400 GeV protons, $\theta_{in}=2\mu rad$, $l_{cr}=0.48 mm$, $\theta_b=170\mu rad$, (110) planes; (e) SLAC, 20.35 GeV positrons, $\theta_{in}=10\mu rad$, $l_{cr}=0.11 mm$, $\theta_b=1600\mu rad$, (110) planes; (f) the same as (b) for (111) planes for positrons.}
 \end{figure}

The program CRYSTAL \cite{Sytov,Sytov2} was applied for simulations. This program was validated in particular in the simulation of single-pass experiments at CERN SPS \cite{H8}, in which a good agreement of experimental and simulation results was achieved \cite{Sytov}. The model implemented in this program \cite{Tikh1,Tikh12,Tikh2,BaryshevskyTikhomirov} was also successfully applied to the explanation of different experiments on channeling radiation \cite{BaryshevskyTikhomirov,TikhPRL,TikhPRL2}. The effect of multiple volume reflection in a single-piece bent crystal was also predicted in the frame of this model \cite{MVROC,MVROCTGM,MVROCTS} and demonstrated in several experiments \cite{MVROCp,MVROCn,MVROCp2}. The code is based on charged particles trajectory Monte Carlo simulation in a continuum potential of crystal planes or axes. Besides, Coulomb single- and multiple-scattering on either nuclei or electrons is simulated. In addition, the simulation of nuclear scattering is implemented. It is important to stress that realistic trajectory simulation ``from the first principles'' without any simplifications and approximations is essential for dechanneling, rechanneling and volume-capture effects. In fact, only realistic simulation allows one to predict the new effects, in particular those, described in this paper.

The simulations were performed for a single passage of charged particles through the crystal. The typical statistics was $10^6$ particles. The angular divergence of the initial beam was set to be less than $\theta_L/4$. The bent crystal parameters were chosen to fulfil the conditions (\ref{2}-\ref{3}, \ref{19}) for clear observation of both channeling and quasichanneling oscillations.

The simulation of the distributions of the particle angles after interaction of particles with the crystal is shown in Figs. \ref{F1}-\ref{F4} for channeling and volume reflection orientations, respectively. The energy in the simulation was chosen in correspondence to the beam energy in currently operating accelerators.

Both channeling and quasichanneling oscillations are observable. Moreover, they are in a good agreement with the estimations obtained above. In particular, the simulated interpeak distance for channeling oscillations is consistent with the estimation of channeling oscillation length (\ref{1}). The highest deviation is for 20.35 GeV because in that case the bending radius is close to the critical radius. The correlations quickly disappear also because of rather small bending radius. As mentioned above for volume reflection, the interpeak distance corresponds to one oscillation length, in contrast to channeling for which interpeak distance is half of one oscillation length. As expected, the planar channeling oscillations are not observed for negative particles.

Simulated quasichanneling oscillations agree with formulae (\ref{13}-\ref{17}) for both signs of particles even for the first oscillation, i.e., the closest one to the channeling peak. Vertical lines calculated by (\ref{13}) define the location of peaks for quasichanneling oscillations and their agreement with the simulations. It is important to underline that for the case of 7 TeV the quasichanneling oscillations are revealed not in the zones predicted by Eq. (\ref{13}) but at intersections of such zones.

It is important to emphasize that the angular difference (\ref{17}-\ref{18}) is on the left of the peak located at $\theta_{Xdefl}$. The angular distance between the peaks decreases w.r.t. the angle measured from the channeling peak. In addition, the particles in neighboring peaks undergo more oscillations and travel longer under over-barrier state, resulting in increased scattering angle. Because of this, only the peaks of quasichanneling oscillations near the channeling bump can be observed. Qualitatively, this is the manifestation of condition (\ref{19}).

The angular distribution of particles after their interaction with the crystal was obtained also with the increased angular divergence of the initial beam by approximately $3/4 \theta_{L}$. The corresponding cases are shown in Figs. \ref{F1} and \ref{F4} for 150 GeV. Indeed, too large an angular divergence leads to the disappearance of the peaks of planar channeling oscillations. As mentioned above, quasichanneling oscillations do not directly depend on the angular divergence. Thereby, such peaks remain visible.

In Fig. \ref{F1} the four upper plots represent the scaling on energy (\ref{6}-\ref{7}) introduced in the previous section. The same scaling is represented in Fig. \ref{F4} for volume reflection.

Such scaling is good for the energies of the same order. In the opposite case, the radius can approach to the critical one, when the conditions for the observation of the planar channeling oscillations are not optimal. This is shown in Fig. \ref{F1} for channeling and in Fig. \ref{F4} for volume reflection. The obtained scaling provides a similar picture for different energies from hundreds of GeV up to 7 TeV. At the same time, the picture for the case of 20.35 GeV is different and not so evident because the bending radius approaches to its critical radius.

For the (111) crystal planes the picture observed is analogous to that for the (110) planes (see Figs. \ref{F1}-\ref{F11}). For planar channeling oscillations the interpeak distance is proportional to the channeling length in the larger channel as shown in Fig. \ref{F2}. The quasichanneling oscillations are well described by formulae (\ref{13}-\ref{17}) if the interplanar distance is determined as a transverse period being equal to $3.13{\AA}$ for silicon.

\section{On the experimental observation of channeling and quasichanneling oscillations}

\begin{figure}
\resizebox{77mm}{!}{\includegraphics{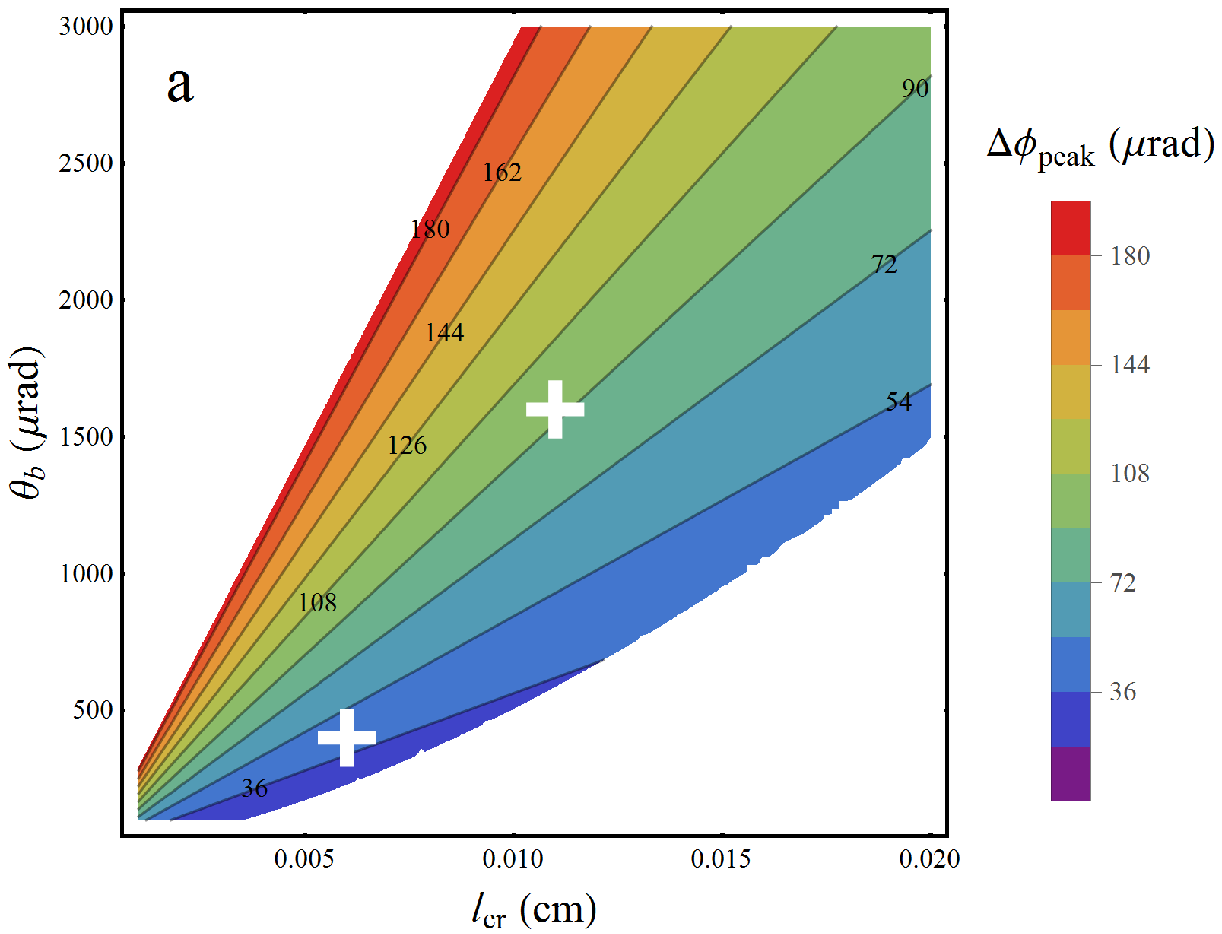}}
\resizebox{77mm}{!}{\includegraphics{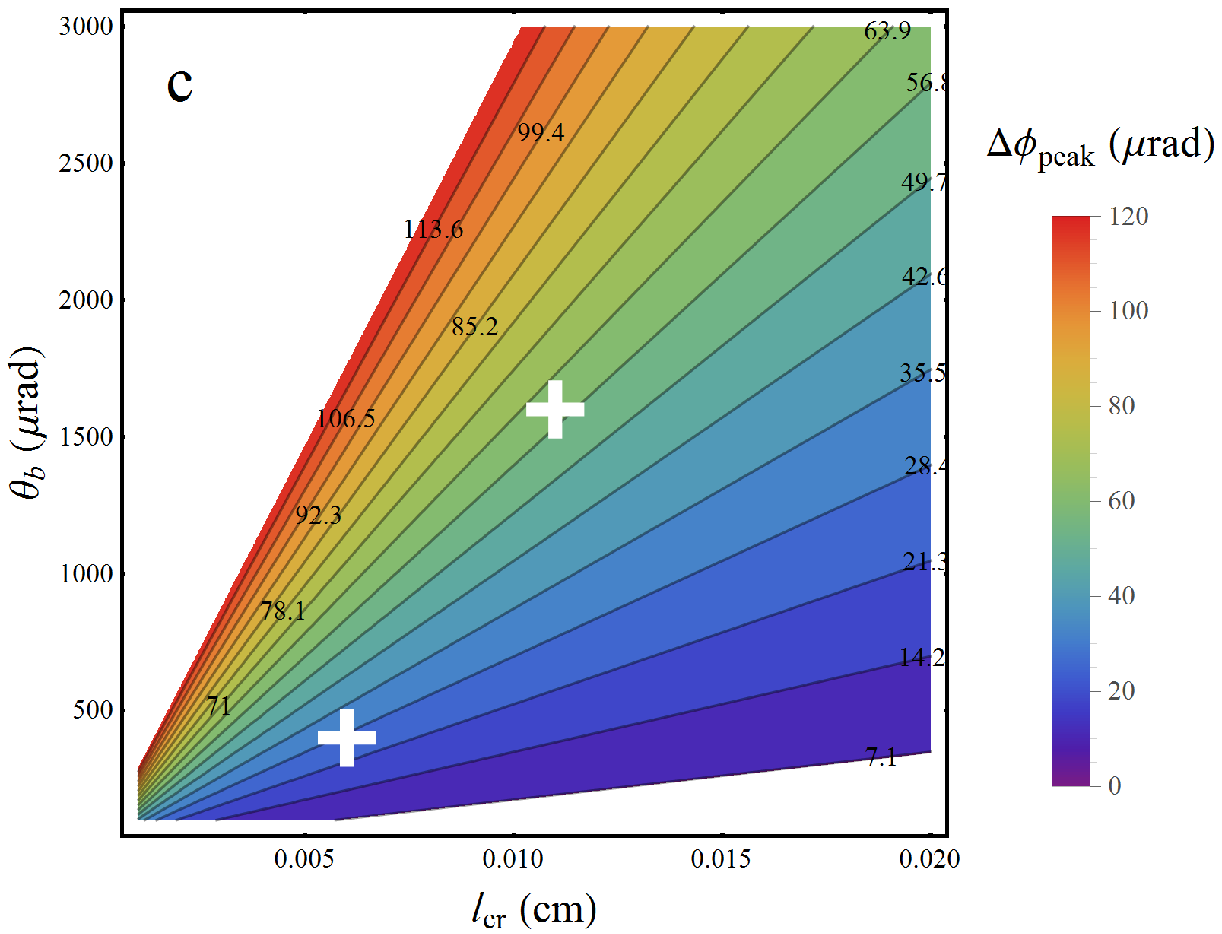}}
\resizebox{77mm}{!}{\includegraphics{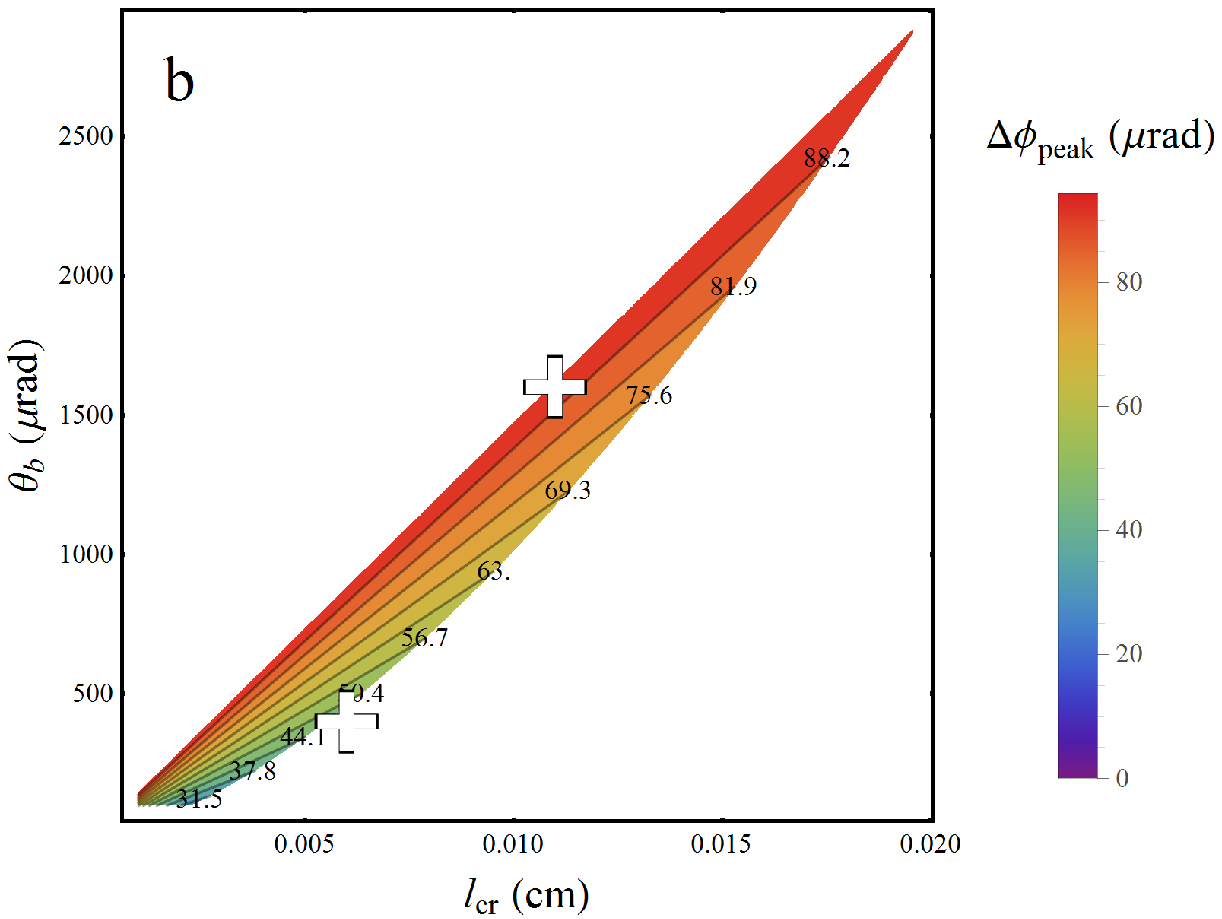}}
\resizebox{77mm}{!}{\includegraphics{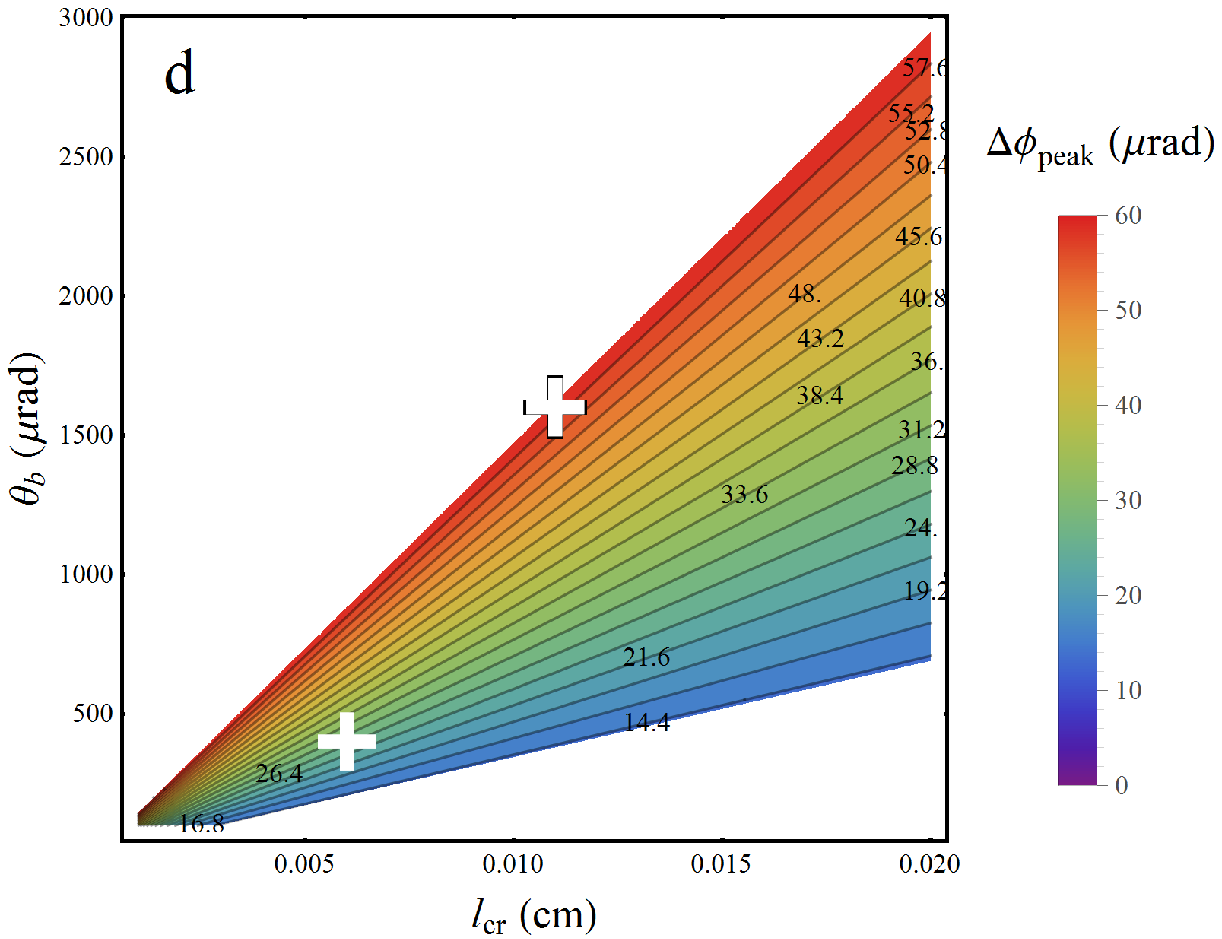}}
 \caption{\label{F5} The dependence of the angular distance between the peaks corresponding to the channeling (a) and quasichanneling (c) oscillations on the crystal length and the bending angle for the positrons of 20.35 GeV. The (110) silicon planes are considered. The zone for the dechanneling peaks observation is formed by (\ref{2}-\ref{3}). (b,d):  the optimal zones for the same cases for the channeling and quasichanneling oscillations respectively, the ratios in (\ref{2}-\ref{3}) and (\ref{19}, \ref{3}) exceed two.}
 \end{figure}

In this section we provide information on possible experimental set-ups for observation of planar channeling and quasichanneling oscillations through existing accelerators worldwide. In order to span over different energy and charge, we considered the cases of both positrons and electrons at SLAC (20.35 GeV) and SPS, CERN (150 GeV) and of electrons at MAMI (855 MeV).

For successful observation, it is very important to choose the proper parameters for the crystal geometry. They are provided by Eqs. (\ref{2}-\ref{3}) for channeling oscillations and by Eqs. (\ref{19}, \ref{3}) for quasichanneling. However, in a real experiment the angular distance between the peaks should be as large as possible to better resolve them. All these conditions can be visually combined in the interpeak dependence of the distance between the peaks on the crystal length and bending angle.

Such dependence is shown in Fig. \ref{F5} for the channeling oscillations at the channeling orientation of the energy of 20.35 GeV. White crosses mark the crystal geometry simulated in this paper and presented in Figs. \ref{F1}-\ref{F11}. The conditions (\ref{2}-\ref{3}) determine the area, where the observation of planar channeling oscillations is allowed. In order to optimize the crystal parameters the ratios (\ref{2}-\ref{3}) should be safely taken as 2-3 times as much. For equation (\ref{2}) this choice results in a clearer picture of the peaks. At the same time, for the ratio (\ref{3}), it provides higher channeling efficiency for better statistics of the experiment.

An example of optimized zone for crystal geometry for a SLAC case is shown in Fig. \ref{F5}b. The estimates (\ref{2}-\ref{3}) provide a sufficiently narrow region of crystal parameters. However, in any case a concrete experimental layout should be checked by Monte-Carlo trajectory simulations.

Similar conclusion can be inferred by application of Eqs. (\ref{19}, \ref{3}) to the plots for quasichanneling oscillations. These dependencies are represented in Fig. \ref{F5} for 20.35 GeV. White crosses in the optimal zone indicate the parameters used in this paper. Quasichanneling oscillations are indeed observed for our simulations for all the cases considered.

The algorithm for crystal geometry optimization remains the same as for the quasichanneling oscillations. The only difference is that the initial angular divergence of the beam should be much less important than for planar channeling oscillations. The angular divergence in our simulations was equal to 10 $\mu rad$, a value which may be experimentally achieved. Thus, the SLAC case satisfies all the conditions of the observation of planar channeling and quasichanneling oscillations.

For the SPS case, the crucial factor is the angular resolution of the detector. At energies of the order of hundreds GeV order, the resolution of at least several microradians should be provided. At the SPS, additional scattering by air and the silicon strip detectors contributes to the measurements. This contribution can be taken into account by including the corresponding r.m.s. scattering angle $\theta_{det}$ to the denominator of (\ref{2}) and (\ref{19}):
\begin{equation}
\label{9}
\frac{\Delta \varphi _{ch}}{2 \sqrt{\theta_{sc}^{2}+\theta_{det}^2}}>1.
\end{equation}
The angular divergence at the SPS is expected to be higher when using secondary beams of positrons or electrons. In this case only quasichanneling oscillations can be observed.

\begin{figure}
\resizebox{89mm}{!}{\includegraphics{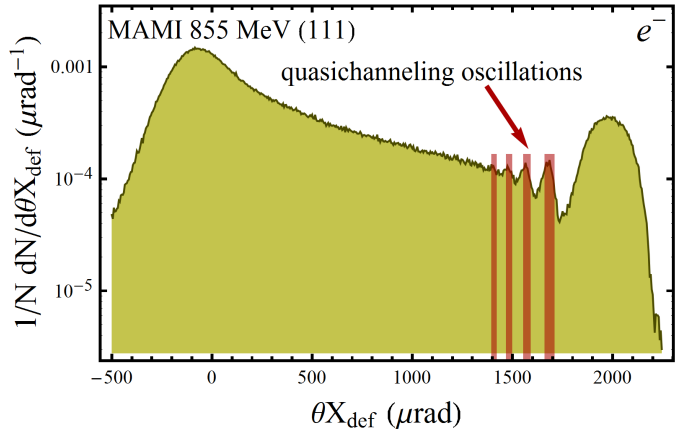}}
 \caption{\label{F9} The angular distributions of particles after interaction with the silicon crystal at the channeling orientation. The simulation layout is: 855 MeV electrons, $\theta_{in}=50\mu rad$, $l_{cr}=15 \mu m$, $\theta_b=2000\mu rad$, (111) planes.}
 \end{figure}

For electrons only quasichanneling oscillations can be observed as at the MAMI microtron \cite{TikhPRL2}. The simulation of such experiment is shown in Fig. \ref{F9}. The main problem for this experiment is crystal manufacturing. For operation of sub-GeV energies, a very short and strongly bent crystal is required, which is at the limit of existing technologies.

(111) crystal planes provide wider angular distance between the peaks. Thus, they should be preferable. Moreover, it is simpler to manufacture (111) bent crystal than for any direction, when strong bending is required \cite{Giacomo}. The latter case especially relates to smaller energies of 1 GeV order at which such crystals were successfully applied \cite{U702,TikhPRL2,SLAC2}. For electrons (111) planes provide a deeper potential well than the (110) ones. This results in a higher channeling efficiency which is also preferred.

\section{Conclusions}

The effect of \textit{planar quasichanneling oscillations} in the deflection angle distribution of particles passed through a bent crystal has been predicted. The effect of planar channeling oscillations was also analyzed. Both of them possess a fine structure in the angular distribution as visualized by Monte Carlo simulations for a wide range of energies.

The theoretical interpretation of both kinds of oscillations was proposed. Quasichanneling oscillations appear near the direction at which channeling particles leave the crystal. They arise due to the correlations of over-barrier oscillation lengths of dechanneled particles. Channeling oscillations can be observed in all over the angular range of deflected particles after interaction with a crystal. This effect arises from correlated dechanneling of particles moving along phase-correlated trajectories under channeling mode. An equation for the angular positions of quasichanneling peaks was found. It demonstrates the independence of peak position on charge sign and energy.

Since phase correlation for channeled particles is conserved only for positive particles, the channeling oscillation peaks can not be observed for negative charges. At the same time, since both negatively and positively charged particles may experience over-barrier oscillations, the effect of quasichanneling oscillations can be observed for both of them.

The possibility to observe both channeling and quasichanneling oscillations is limited by incoherent scattering of particles under over-barrier states. Both of them can be observed if only the r.m.s. angle of incoherent scattering is twice smaller than the interpeak angular intervals. The angular resolution of particle detectors is crucial for the observation of both types of oscillations. However, the low angular divergence of the incident beam is necessary only for an observation of the channeling oscillations.

The optimal conditions for experimental observation of both channeling and quasichanneling oscillations are also proposed. These conditions are applied to elaborate the optimal values of crystal thickness and bending angle (radius) at SLAC, SPS, MAMI and LHC. A comparison of (110) and (111) planar crystal orientation reveals the higher interpeak distance and higher electron channeling efficiency in the case of the latter. (111) orientation is also preferable from the point of view of strong bending of thin crystals to observe the predicted effects at the SLAC and MAMI energies.

Similarly to channeling oscillations, which are used in low-energy RBS experiments to assess the quality of a crystal, channeling and quasichanneling oscillations could be used to determine the precision of alignment of a high-energy beam with a crystal. In fact, the pattern of the distribution of particles after interaction with a bent crystal is highly sensitive to the beam-to-crystal alignment. This information can be used for all the applications for which bent crystals are used in accelerators, such as beam collimation, extraction and e.m. radiation generation.

\section{Acknowledgements}

We acknowledge partial support by INFN under the CHANEL experiment. This work is also supported by the Belarusian Foundation for Basic Research Grants and the Ministry of Education of the Republic of Belarus under contract No. F14MV-010.

\end{document}